\documentclass[aps,prl,twocolumn,groupedaddress,notitlepage,
floatfix,superscriptaddress]{revtex4-1}

\usepackage{graphicx,graphics,epsfig,subfigure,times,bm,bbm,amssymb,amsmath,amsthm,mathrsfs,MnSymbol}
\usepackage{gensymb}
\usepackage{amsfonts}
\usepackage{float}
\usepackage[matrix,frame,arrow]{xypic}
\usepackage[pdfstartview=FitH]{hyperref}
\hypersetup{
	colorlinks=true,       
	linkcolor=red,          
	citecolor=blue,        
	filecolor=magenta,      
	urlcolor=blue,           
	runcolor=cyan
}
\usepackage[pdftex]{color}
\usepackage{braket}  
\usepackage{enumerate}
\usepackage[normalem]{ulem}
\usepackage[usenames,dvipsnames]{xcolor}
\usepackage{multirow}
\usepackage{mathtools}
\usepackage{bbm}
\usepackage{titletoc}

\definecolor{orange}{rgb}{1,0.5,0}

\newcommand{\RNum}[1]{\uppercase\expandafter{\romannumeral #1\relax}}

\newcommand{\ignore}[1]{}
\usepackage{geometry}

\ignore{
	\documentclass[eprintnumbers,amsmath,amssymb,onecolumn,a4paper,caption ]{article}
	\usepackage{amsfonts}
	\usepackage{amssymb}
	\usepackage{mathrsfs}
	\usepackage{mathbbold}
	\usepackage{bbm}
	\usepackage{mathrsfs}
	\usepackage{dcolumn}
	\usepackage{bm}
	\usepackage{times,epsfig,amssymb,amsmath}
	\usepackage{float}
	\usepackage{subfigure}
	\usepackage{geometry}\geometry{left=2.5cm,right=2.5cm,top=3cm,bottom=3cm}

	\usepackage{color}

}
\begin{document}   
	
	\title{Realization of high-fidelity CZ gates in extensible superconducting qubits design with a tunable coupler}

	\author{Yangsen~Ye}
	\affiliation{Hefei National Laboratory for Physical Sciences at the Microscale and Department of Modern Physics, University of Science and Technology of China, Hefei 230026, China}
	\affiliation{Shanghai Branch, CAS Center for Excellence in Quantum Information and Quantum Physics, University of Science and Technology of China, Shanghai 201315, China}
	\affiliation{Shanghai Research Center for Quantum Sciences, Shanghai 201315, China}

	\author{Sirui~Cao}
	\affiliation{Hefei National Laboratory for Physical Sciences at the Microscale and Department of Modern Physics, University of Science and Technology of China, Hefei 230026, China}
	\affiliation{Shanghai Branch, CAS Center for Excellence in Quantum Information and Quantum Physics, University of Science and Technology of China, Shanghai 201315, China}
	\affiliation{Shanghai Research Center for Quantum Sciences, Shanghai 201315, China}
	
	\author{Yulin~Wu}
	\affiliation{Hefei National Laboratory for Physical Sciences at the Microscale and Department of Modern Physics, University of Science and Technology of China, Hefei 230026, China}
	\affiliation{Shanghai Branch, CAS Center for Excellence in Quantum Information and Quantum Physics, University of Science and Technology of China, Shanghai 201315, China}
	\affiliation{Shanghai Research Center for Quantum Sciences, Shanghai 201315, China}
	
	\author{Xiawei~Chen}
	\affiliation{Shanghai Branch, CAS Center for Excellence in Quantum Information and Quantum Physics, University of Science and Technology of China, Shanghai 201315, China}
	
	\author{Qingling~Zhu}
	\affiliation{Hefei National Laboratory for Physical Sciences at the Microscale and Department of Modern Physics, University of Science and Technology of China, Hefei 230026, China}
	\affiliation{Shanghai Branch, CAS Center for Excellence in Quantum Information and Quantum Physics, University of Science and Technology of China, Shanghai 201315, China}
	\affiliation{Shanghai Research Center for Quantum Sciences, Shanghai 201315, China}
	
	\author{Shaowei~Li}
	\affiliation{Hefei National Laboratory for Physical Sciences at the Microscale and Department of Modern Physics, University of Science and Technology of China, Hefei 230026, China}
	\affiliation{Shanghai Branch, CAS Center for Excellence in Quantum Information and Quantum Physics, University of Science and Technology of China, Shanghai 201315, China}
	\affiliation{Shanghai Research Center for Quantum Sciences, Shanghai 201315, China}
	
	\author{Fusheng~Chen}
	\affiliation{Hefei National Laboratory for Physical Sciences at the Microscale and Department of Modern Physics, University of Science and Technology of China, Hefei 230026, China}
	\affiliation{Shanghai Branch, CAS Center for Excellence in Quantum Information and Quantum Physics, University of Science and Technology of China, Shanghai 201315, China}
	\affiliation{Shanghai Research Center for Quantum Sciences, Shanghai 201315, China}
	
	\author{Ming~Gong}
	\affiliation{Hefei National Laboratory for Physical Sciences at the Microscale and Department of Modern Physics, University of Science and Technology of China, Hefei 230026, China}
	\affiliation{Shanghai Branch, CAS Center for Excellence in Quantum Information and Quantum Physics, University of Science and Technology of China, Shanghai 201315, China}
	\affiliation{Shanghai Research Center for Quantum Sciences, Shanghai 201315, China}
	
	\author{Chen~Zha}
	\affiliation{Hefei National Laboratory for Physical Sciences at the Microscale and Department of Modern Physics, University of Science and Technology of China, Hefei 230026, China}
	\affiliation{Shanghai Branch, CAS Center for Excellence in Quantum Information and Quantum Physics, University of Science and Technology of China, Shanghai 201315, China}
	\affiliation{Shanghai Research Center for Quantum Sciences, Shanghai 201315, China}
	
	\author{He-Liang Huang}
	\affiliation{Hefei National Laboratory for Physical Sciences at the Microscale and Department of Modern Physics, University of Science and Technology of China, Hefei 230026, China}
	\affiliation{Shanghai Branch, CAS Center for Excellence in Quantum Information and Quantum Physics, University of Science and Technology of China, Shanghai 201315, China}
	\affiliation{Shanghai Research Center for Quantum Sciences, Shanghai 201315, China}
	\affiliation{Henan Key Laboratory of Quantum Information and Cryptography, Zhengzhou 450000, China}
	
	\author{Youwei~Zhao}
	\affiliation{Hefei National Laboratory for Physical Sciences at the Microscale and Department of Modern Physics, University of Science and Technology of China, Hefei 230026, China}
	\affiliation{Shanghai Branch, CAS Center for Excellence in Quantum Information and Quantum Physics, University of Science and Technology of China, Shanghai 201315, China}
	\affiliation{Shanghai Research Center for Quantum Sciences, Shanghai 201315, China}
	
	\author{Shiyu~Wang}
	\affiliation{Hefei National Laboratory for Physical Sciences at the Microscale and Department of Modern Physics, University of Science and Technology of China, Hefei 230026, China}
	\affiliation{Shanghai Branch, CAS Center for Excellence in Quantum Information and Quantum Physics, University of Science and Technology of China, Shanghai 201315, China}
	\affiliation{Shanghai Research Center for Quantum Sciences, Shanghai 201315, China}
	
	\author{Shaojun~Guo}
	\affiliation{Hefei National Laboratory for Physical Sciences at the Microscale and Department of Modern Physics, University of Science and Technology of China, Hefei 230026, China}
	\affiliation{Shanghai Branch, CAS Center for Excellence in Quantum Information and Quantum Physics, University of Science and Technology of China, Shanghai 201315, China}
	\affiliation{Shanghai Research Center for Quantum Sciences, Shanghai 201315, China}
	
	\author{Haoran~Qian}
	\affiliation{Hefei National Laboratory for Physical Sciences at the Microscale and Department of Modern Physics, University of Science and Technology of China, Hefei 230026, China}
	\affiliation{Shanghai Branch, CAS Center for Excellence in Quantum Information and Quantum Physics, University of Science and Technology of China, Shanghai 201315, China}
	\affiliation{Shanghai Research Center for Quantum Sciences, Shanghai 201315, China}
	
	\author{Futian~Liang}
	\affiliation{Hefei National Laboratory for Physical Sciences at the Microscale and Department of Modern Physics, University of Science and Technology of China, Hefei 230026, China}
	\affiliation{Shanghai Branch, CAS Center for Excellence in Quantum Information and Quantum Physics, University of Science and Technology of China, Shanghai 201315, China}
	\affiliation{Shanghai Research Center for Quantum Sciences, Shanghai 201315, China}
	
	\author{Jin~Lin}
	\affiliation{Hefei National Laboratory for Physical Sciences at the Microscale and Department of Modern Physics, University of Science and Technology of China, Hefei 230026, China}
	\affiliation{Shanghai Branch, CAS Center for Excellence in Quantum Information and Quantum Physics, University of Science and Technology of China, Shanghai 201315, China}
	\affiliation{Shanghai Research Center for Quantum Sciences, Shanghai 201315, China}
	
	\author{Yu~Xu}
	\affiliation{Hefei National Laboratory for Physical Sciences at the Microscale and Department of Modern Physics, University of Science and Technology of China, Hefei 230026, China}
	\affiliation{Shanghai Branch, CAS Center for Excellence in Quantum Information and Quantum Physics, University of Science and Technology of China, Shanghai 201315, China}
	\affiliation{Shanghai Research Center for Quantum Sciences, Shanghai 201315, China}
	
	\author{Cheng~Guo}
	\affiliation{Hefei National Laboratory for Physical Sciences at the Microscale and Department of Modern Physics, University of Science and Technology of China, Hefei 230026, China}
	\affiliation{Shanghai Branch, CAS Center for Excellence in Quantum Information and Quantum Physics, University of Science and Technology of China, Shanghai 201315, China}
	\affiliation{Shanghai Research Center for Quantum Sciences, Shanghai 201315, China}
	
	\author{Lihua~Sun}
	\affiliation{Hefei National Laboratory for Physical Sciences at the Microscale and Department of Modern Physics, University of Science and Technology of China, Hefei 230026, China}
	\affiliation{Shanghai Branch, CAS Center for Excellence in Quantum Information and Quantum Physics, University of Science and Technology of China, Shanghai 201315, China}
	\affiliation{Shanghai Research Center for Quantum Sciences, Shanghai 201315, China}
	
	\author{Na~Li}
	\affiliation{Hefei National Laboratory for Physical Sciences at the Microscale and Department of Modern Physics, University of Science and Technology of China, Hefei 230026, China}
	\affiliation{Shanghai Branch, CAS Center for Excellence in Quantum Information and Quantum Physics, University of Science and Technology of China, Shanghai 201315, China}
	\affiliation{Shanghai Research Center for Quantum Sciences, Shanghai 201315, China}

	\author{Hui~Deng}
	\affiliation{Hefei National Laboratory for Physical Sciences at the Microscale and Department of Modern Physics, University of Science and Technology of China, Hefei 230026, China}
	\affiliation{Shanghai Branch, CAS Center for Excellence in Quantum Information and Quantum Physics, University of Science and Technology of China, Shanghai 201315, China}
	\affiliation{Shanghai Research Center for Quantum Sciences, Shanghai 201315, China}
	
	\author{Xiaobo~Zhu}
	\email{xbzhu16@ustc.edu.cn}
	\affiliation{Hefei National Laboratory for Physical Sciences at the Microscale and Department of Modern Physics, University of Science and Technology of China, Hefei 230026, China}
	\affiliation{Shanghai Branch, CAS Center for Excellence in Quantum Information and Quantum Physics, University of Science and Technology of China, Shanghai 201315, China}
	\affiliation{Shanghai Research Center for Quantum Sciences, Shanghai 201315, China}
	
	\author{Jian-Wei~Pan}
	\affiliation{Hefei National Laboratory for Physical Sciences at the Microscale and Department of Modern Physics, University of Science and Technology of China, Hefei 230026, China}
	\affiliation{Shanghai Branch, CAS Center for Excellence in Quantum Information and Quantum Physics, University of Science and Technology of China, Shanghai 201315, China}
	\affiliation{Shanghai Research Center for Quantum Sciences, Shanghai 201315, China}

	\begin{abstract}
		High-fidelity two-qubits gates are essential for the realization of large-scale quantum computation and simulation. Tunable coupler design is used to reduce the problem of parasitic coupling and frequency crowding in many-qubit systems and thus thought to be advantageous. Here we design a extensible 5-qubit system in which center transmon qubit can couple to every four near-neighbor qubit via a capacitive tunable coupler and experimentally demonstrate high-fidelity controlled-phase (CZ) gate by manipulating center qubit and one near-neighbor qubit. Speckle purity benchmarking (SPB) and cross entrophy benchmarking (XEB) are used to assess the purity fidelity and the fidelity of the CZ gate. The average purity fidelity of the CZ gate is 99.69$\pm$0.04\% and the average fidelity of the CZ gate is 99.65$\pm$0.04\% which means the control error is about 0.04\%. Our work will help resovle many chanllenges in the implementation of large scale quantum systems. 
	\end{abstract}
	
	\maketitle 
	
	
	Over the past 20 years, superconducting qubits have made great progress in both quantity and quality~\cite{Krantz}. With the increase of the number of qubits and the improvement of the gate fidelity, superconducting circuits have emerged as a powerful platform of quantum simulation~\cite{Buluta,Houck,Georgescu} and also as a promising implementation for fault-tolerant quantum computation~\cite{Devoret,Campbell}. For typical superconducting transmon/Xmon qubits~\cite{Koch2007,Barends}, there are many proposals to realize two-qubits gates. The main kind of proposals is implemented with frequency-tunable qubits with static capacitive couplings. By carefully tuning the frequency of qubits to make the $|11\rangle$ state in resonance with the $|02\rangle$ state, CZ gate with low leakage to high energy levels has been implemented~\cite{Martinis}. But due to the static coupling, two qubits’ idle frequencies should be set far away from each other to reduce the residual $ZZ$ coupling between $|11\rangle$ and $|02\rangle$ and lower the error when simultaneously performing single qubit gates operation. When the number of qubits of the system increases, the frequency crowding problem becomes worse. Superconducting qubits with a tunable coupler~\cite{Chen2014,YanF2018,Niskanen,Hime} have been studied as a proposal to sovle these problems and high-fidelity two-qubits gates have been implemented experimentally~\cite{sung2020,Foxen}.  
	
	In this work, we experimentally realize a extensible superconducting circuits with four tunable couplers and five transmon qubits and implement high-fidelity two-qubit CZ gate between the center qubit and one of the four near-neighbor qubits. Via optimized control, we demonstrate a two-qubit CZ gate with average fidelity of 99.65\% in cross entrophy benchmarking~\cite{DgateXEB2019,XEB2018}. We also use speckle purity benchmarking (SPB)~\cite{Google-supermacy2019} to assess the average purity fidelity which is 99.69\% of the CZ gate and get the control error is 0.04\%. This result means that we can expand the design to two-dimension structure and realize two-qubit CZ gates between any qubit and one of its near-neighbor qubits. The design in this work and the proposals to realize CZ gate may pave the way to realize fault-tolerant quantum computation.
	
	

	\begin{figure}[h]
	\centering
	\includegraphics[width=8.2cm]{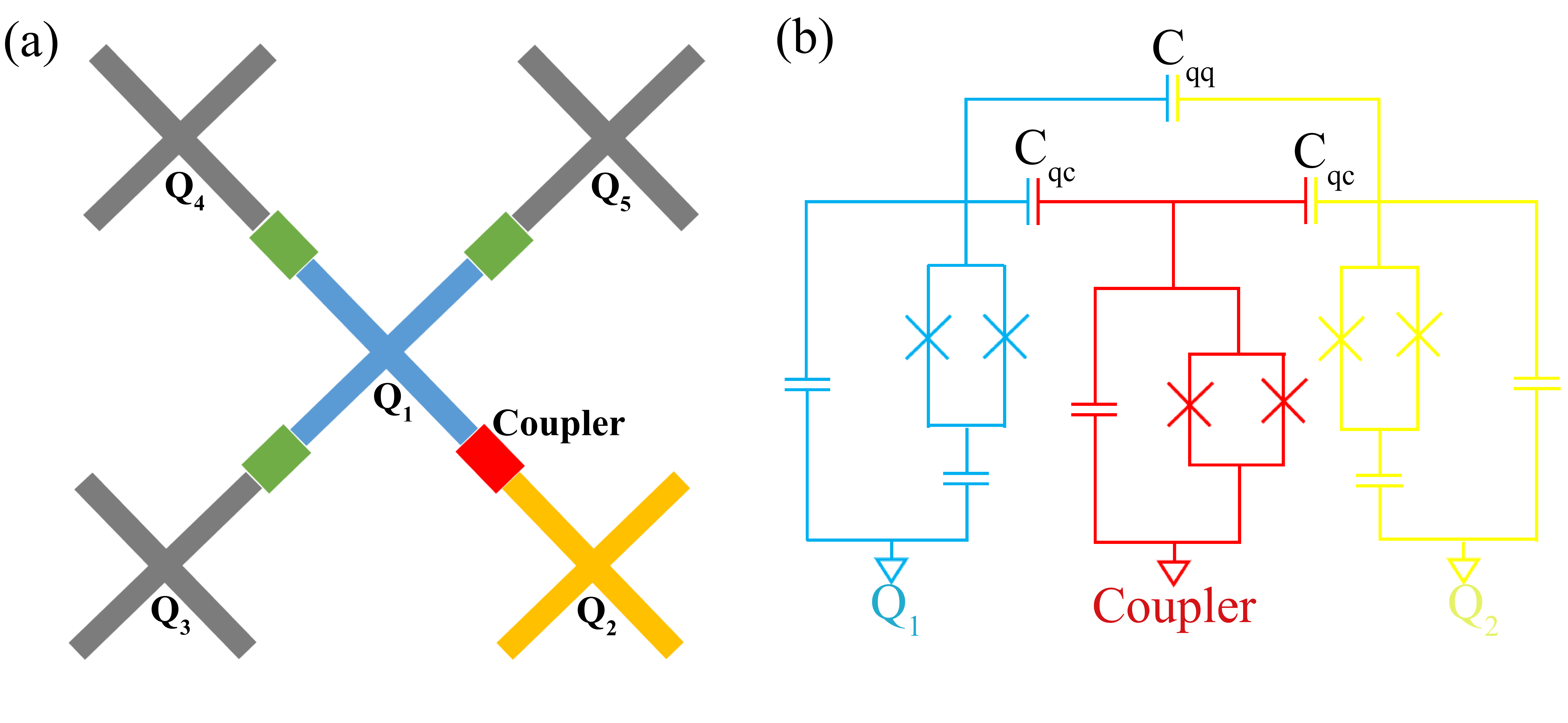}
	\caption{ (a) Design of the qubit chip of the device. There are five transmon qubits ($\text{Q}_{1}$-$\text{Q}_{5}$) and four tunable couplers. The center qubit ($\text{Q}_{1}$) capacitively couples to the other four qubits. Every nearest neighbours share one coupler. Control lines and readout resonators are in the control chip which is not shown here. (b) Schematic diagram of the $\text{Q}_{1}$,  $\text{Q}_{2}$ and their coupler in the qubit chip. $\text{C}_{qc}$ represents the capacitance between the coupler and the qubit and $\text{C}_{qq}$ represents the direct capacitance between the nearset qubits. Each qubit and coupler has a SQUID for individual Z control. In (a) false colors (green, yellow and red) are used to represent the corresponding components in (b).}
	\label{image}
	\end{figure}
		
	To realize extensible design of two dimension structure, flip chip step has been applied in our device and two chips, one is qubit chip for qubit and coupler capacitive structure and the other is control chip for qubit readout and qubit/coupler control, are fabricated. The design of the qubit chip of the device and the schemetic diagram of the $\text{Q}_{1}$,$\text{Q}_{2}$ and their coupler are shown in Fig.~\ref{image}. There are 5 qubits and 4 couplers in the qubit chip. Each coupler has a fast Z bias control line in the control chip to implement Z control. The qubit frequencies can be tuned to a large range of several GHz by both a dc Z bias control line and a fast Z bias control line. Each qubit also has an inductive XY control line to implement single qubit rotation. The readout resonators are seperated into two groups, and one group of resonators for $\text{Q}_{3}$ and $\text{Q}_{4}$ shares one readout transmission line and the other group of resonators for $\text{Q}_{1}$, $\text{Q}_{2}$ and $\text{Q}_{5}$ shares one purcell filter. Both groups of readout resonators have their signals amplified by Josephson parametric amplifiers (JPA)~\cite{JPA2014}.More experimental details of our device and system are presented in Appendix A and Appendix B.
		
	In our devie, the center qubit $\text{Q}_{1}$ can tunably couple to the other four qubits through direct capacitive coupling and indirect coupling via corresponding coupler. We can implement many quantum simulation proposals and even some fault-tolerant quantum proposals if we can realize high-fidelity qubit gates in this design and thus expand our design. So we focus our attention on $\text{Q}_{1}$,$\text{Q}_{2}$ and their coupler. The other three qubits ($\text{Q}_{3}$, $\text{Q}_{4}$, $\text{Q}_{5}$) are always idled with frequencies below 4 GHz with their dc Z bias control lines and the corresponding couplers are idled at their symmetric point before the experiment start and during the experiment with their fast Z bias control lines which we calls workbias mode. This procedure lower the unwanted effective coupling strength between $\text{Q}_{3}$, $\text{Q}_{4}$, $\text{Q}_{5}$ and $\text{Q}_{1}$ and lower the influence to the readout of $\text{Q}_{1}$. After all these preparatory work, we can write an effective Hamiltonian of the three-body system as ($\hbar=1$)
	\begin{equation}
		\label{Hbh} 
		H=\sum_{i=\text{1,2,c}}(w_{i}\hat{b}_{i}^\dagger \hat{b}_{i}+\frac{\eta_{i}}{2}\hat{b}_{i}^\dagger\hat{b}_{i}^\dagger \hat{b}_{i} \hat{b}_{i})\
		+\sum_{{i}\textless{j}}g_{ij}(\hat{b}_{i}^\dagger \hat{b}_{j}+\hat{b}_{j}^\dagger \hat{b}_{i}),
	\end{equation}
	where $\hat{b}_{i}^\dagger$ and $\hat{b}_{i}$ ($\text{i=1,2,c}$) are the raising and lowering operators of the correponding oscillators. The anharmonicity and energy levels of each oscillators are donated by $\eta_{i}$ and $w_{i}$. 
	
	\begin{figure}[h]
		\centering
		\includegraphics[width=\linewidth]{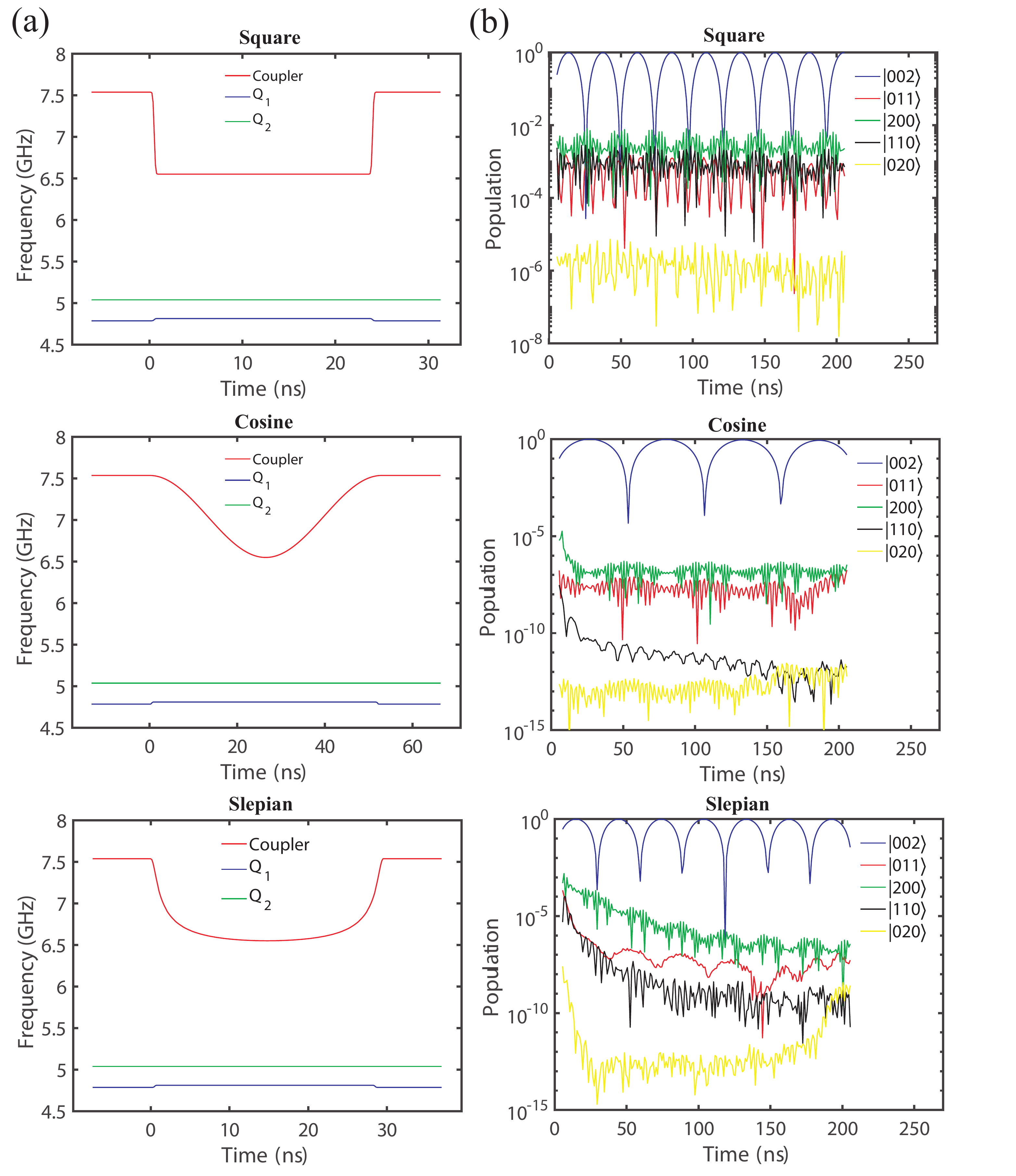}
		\caption{ (a) Waveforms correspond to the highest fidelities for three types of control waveforms. $\text{Q}_{2}$ is placed at the frequency from $|1\rangle$ to $|2\rangle$ of $\text{Q}_{1}$ when idle. (b) Leakage to unwanted energy levels of three types of waveforms.}
		\label{leakage}
	\end{figure}
	
	
	We use the notation $|\text{Q}_{1},\text{Coupler},\text{Q}_{2}\rangle$ to represent the eigenstates of the system (Eq.~\eqref{Hbh}) where Coupler is placed at the frequency that the effective $\text{Q}_{1}-\text{Q}_{2}$ coupling strength is nearly zero. To realize high-fidelity CZ gates, the whole Hamiltonian needs to be considered, especially the existence of coupler. We perform numerical simulations to analyze the three-body system through $\text{QuTiP}$~\cite{QuTiP1,QuTiP2}. Three types of control waveforms of coupler energy level are used, including square-shaped, Slepian-shaped~\cite{Martinis} and cosine-shaped control pulses. For simplification, we do not consider the decoherence of the system and assume the coupling strengths $g_{ij}$ stay the same when energy levels change. We analyze the relationship among the fidelities of the CZ gates, different pulse lengths and the energy level of $\text{Q}_{1}$ of different control waveforms with the same lowest coupler energy level $w_{c}$. In Fig.~\ref{leakage}(a), waveforms which correspond to the highest fidelities of CZ gates of three types are plotted. The pulse lengths needed are different, where square-shaped waveform needs the shortest length around 25 ns, Slepian-shaped waveform needs about 30 ns pulse length and cosine-shaped waveform need the longest pulse length around 63 ns. The highest fidelities of CZ gates of three types of waveforms are also different, where square-shaped waveform's high fidelity is about 99.4$\%$ and the other two waveforms' fidelities can reach above 99.9$\%$. From the fidelities and corresponding pulse lengths, same waveforms with different pulse lengths are used to analyze the leakage to unwanted energy levels in Fig.~\ref{leakage}(b). The square-shaped waveform has the highest periodic leakage up to 0.1$\%$ level which we think is the reason why the CZ gate fidelity of this type of waveform is lower. The other two kinds of waveforms' leakage gradually decreases with the increase of the pulse lengths and can be lower than 0.01$\%$ when pulse lengths are longer than 40ns. Due to the decoherence, we finally choose Slepian-shaped waveform as our experiment waveform.

\begin{figure}[h]
		\centering
		\includegraphics[width=\linewidth]{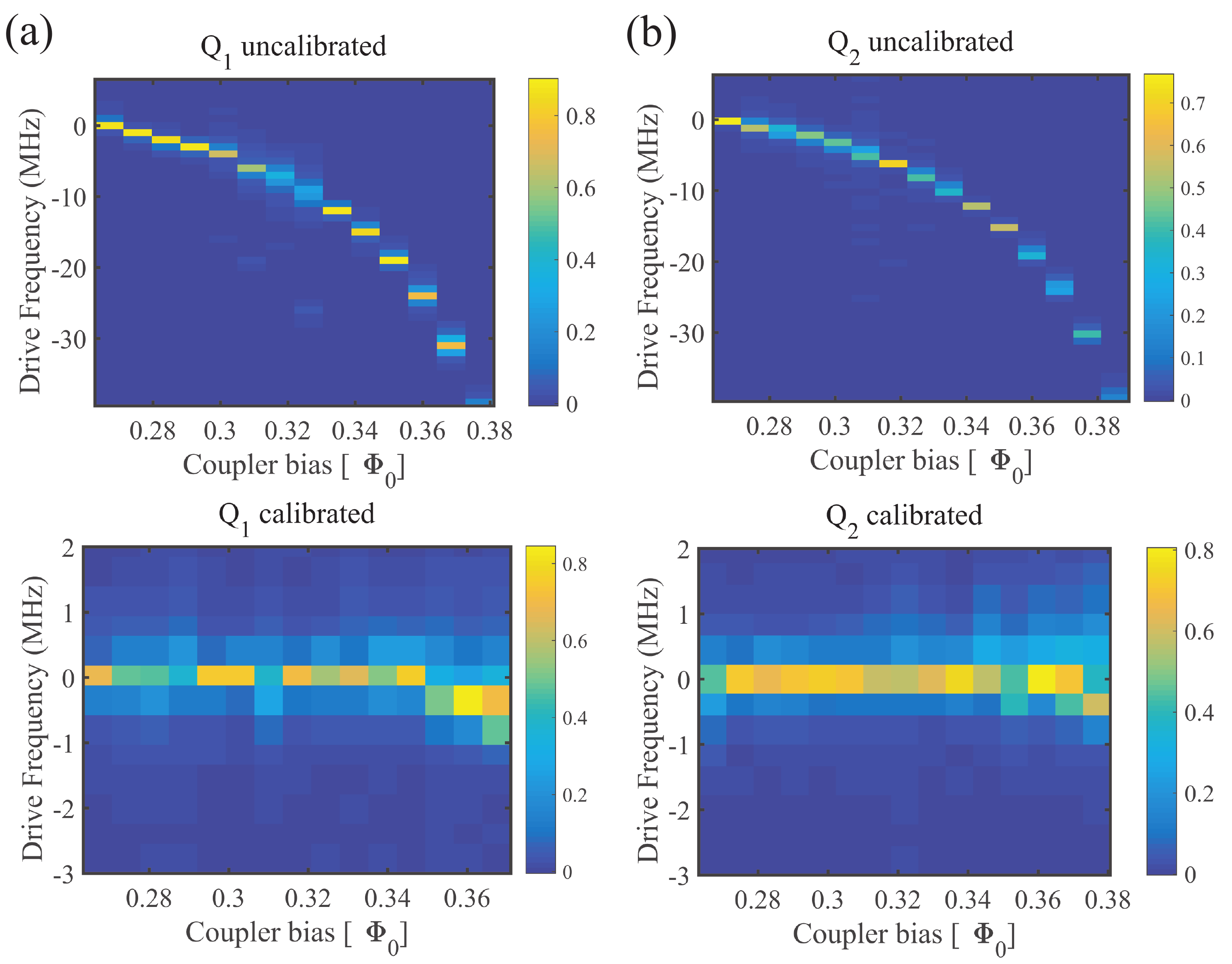}
		\caption{ (a) The frequency of $\text{Q}_{1}$ as a function of the coupler flux bias before and after calibration while $\text{Q}_{2}$ is far detuned. We compensate the frequency shift which ranges from 0 to -30 MHz by tunning the flux bias of $\text{Q}_{1}$, and then sweep the microwave drive frequency and measure the qubit excited state probability. The drive frequency range is based on the original qubit frequency where coupler is placed at the frequency that the effective coupling strength between $\text{Q}_{1}$ and $\text{Q}_{2}$ is nearly zero. (b)  The frequency of $\text{Q}_{2}$ as a function of the coupler flux bias before and after calibration while $\text{Q}_{1}$ is far detuned.}
		\label{calif01}
	\end{figure}

	In our experiment, we have several different points from the numerical simulation. The first difference is that we compensate the frequencies shift of qubits when coupler bias changes. Actually, to avoid two-level-systems, frequencies of qubits when performing CZ gates need to be tunned precisely, especially in many-body systems. The compensation is achieved by first measuring the qubit freuency as a function of the qubit flux bias and then as a function of coupler bias. The uncalibrated and calibrated frequencies of qubits via coupler flux bias are shown in Fig.~\ref{calif01}. The frequencies are almost completely independent on the coupler bias.
	The second difference is that the fast Z bias pulses are distorted when reaching the qubits. This mismatch are corrected by performing deconvolution to the ideal pulse sample data~\cite{Yan}. The third difference is that dc control is not used for couplers, $\text{Q}_{1}$ and $\text{Q}_{2}$ which means their frequencies are not always placed at the idle frequencies. In our experiments, coupler and qubits needed are placed at their original frequencies with zero flux bias at most time and then detuned to the idle frequencies several microseconds before and during the experiment with their fast Z bias control lines and finally detuned back to their original frequencies after the readout pulses end. This control method can reduce the number of control lines in large systems and thus be useful. We can not directly measure the coupler frequency as a function of the coupler flux bias due to the lack of readout resonator of the coupler. So the fourth difference is that we simply use the Slepian-shaped pulse. Relationship between $XX$ coupling strength and coupler flux bias are tested and then fast z control in our experiment is expressed in the form of coupling strength. The last differnece is that $\text{Q}_{2}$'s idle frequency is set around 50 MHz away from the energy level form $|1\rangle$ to $|2\rangle$ of $\text{Q}_{1}$.
	
	\begin{figure}[b]
	\centering
	\includegraphics[width=\linewidth]{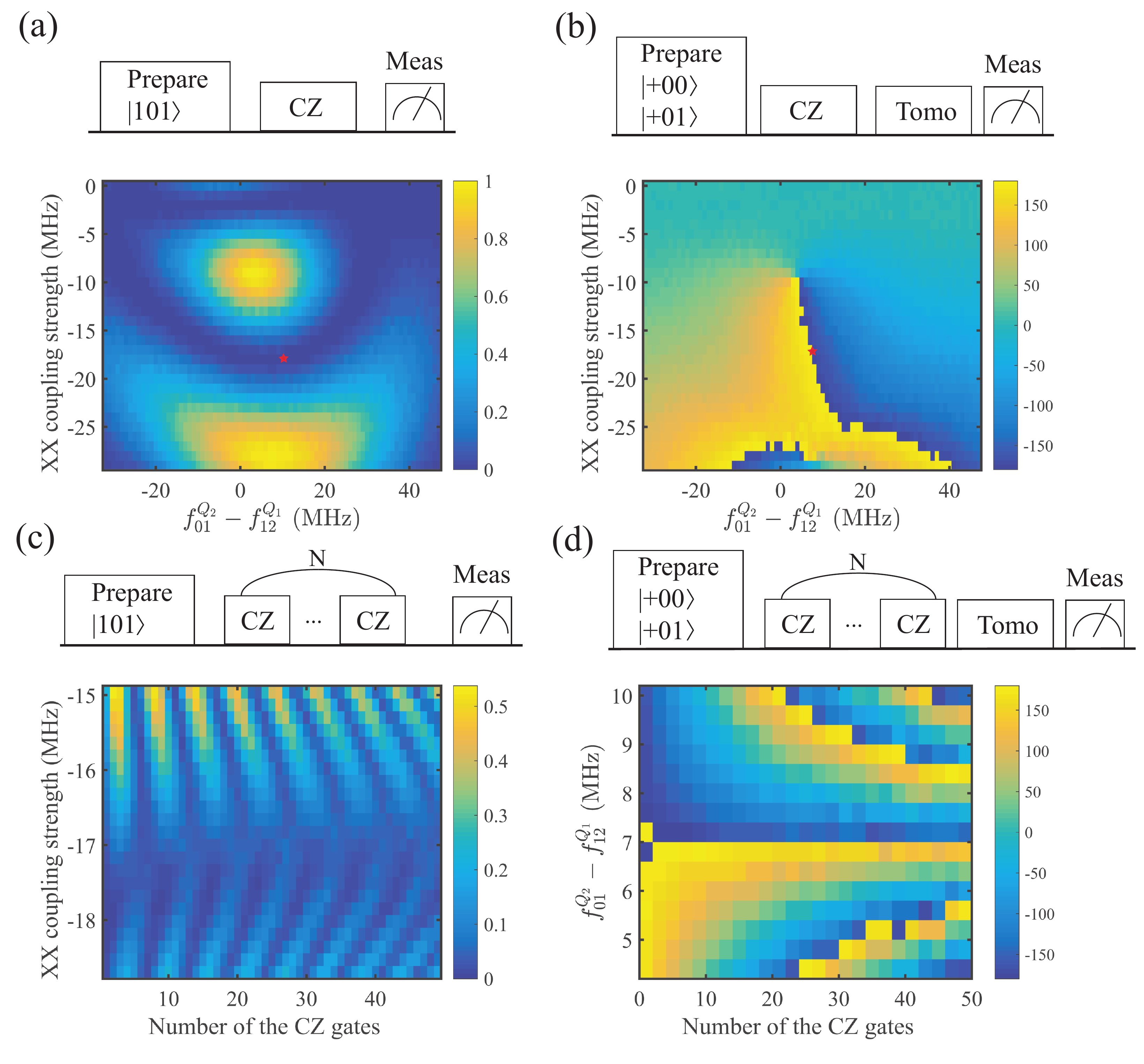}
	\caption{ (a) Schematic of measuring leakage from $|101\rangle$ and the experimental data as a function of $XX$ coupling strength and $\text{Q}_{2}$ detune frequency. (b)  Schematic of a Ramsey-type experiment measuring the conditional phase and the experimental data. Red star in (a) and (b) represents rough CZ point. (c) Delicate measurement of the leakage as a function of $XX$ coupling strength and the number of CZ gates. (d) Delicate measurement of the conditional phase angle ($\phi_{CZ}-(N_{CZ}-1)\times180\degree$) as a function of $\text{Q}_{2}$ detune frequency and the number of CZ gates.  }
	\label{cphase}
	\end{figure}	

	We calibrate the CZ gate by adjusting the Z control amplitudes for a fixed gate length (45 ns) and measuring the conditional phase angle and the leakage from $|101\rangle$. To measure the leakage from $|101\rangle$, we first perform $X$ gates for both two qubits and then measure the state population of $|2i0\rangle$ ($i=\text{0,1,2}$) after a CZ gate (Fig.~\ref{cphase}(a)) since we can not measure the state of the coupler. To measure the conditional phase angle, we perform a Ramsey-type experiment in Fig.~\ref{cphase}(b). The red star in Fig.~\ref{cphase}(a) and (b) represents rough optimal point for the CZ gate which has both low leakage and accurate conditional phase angle. To get more delicate coupler flux bias, we fixed $\text{Q}_{2}$ detune frequency and measure the leakage as a function of $XX$ coupling strength and the number of CZ gates in Fig.~\ref{cphase}(c). After the optimal $XX$ coupling strength decides, we measure the conditional phase angle as a function of $\text{Q}_{2}$ detune frequency and the number of CZ gates in Fig.~\ref{cphase}(d). 
	
\begin{figure}[h]
		\centering
		\includegraphics[width=\linewidth]{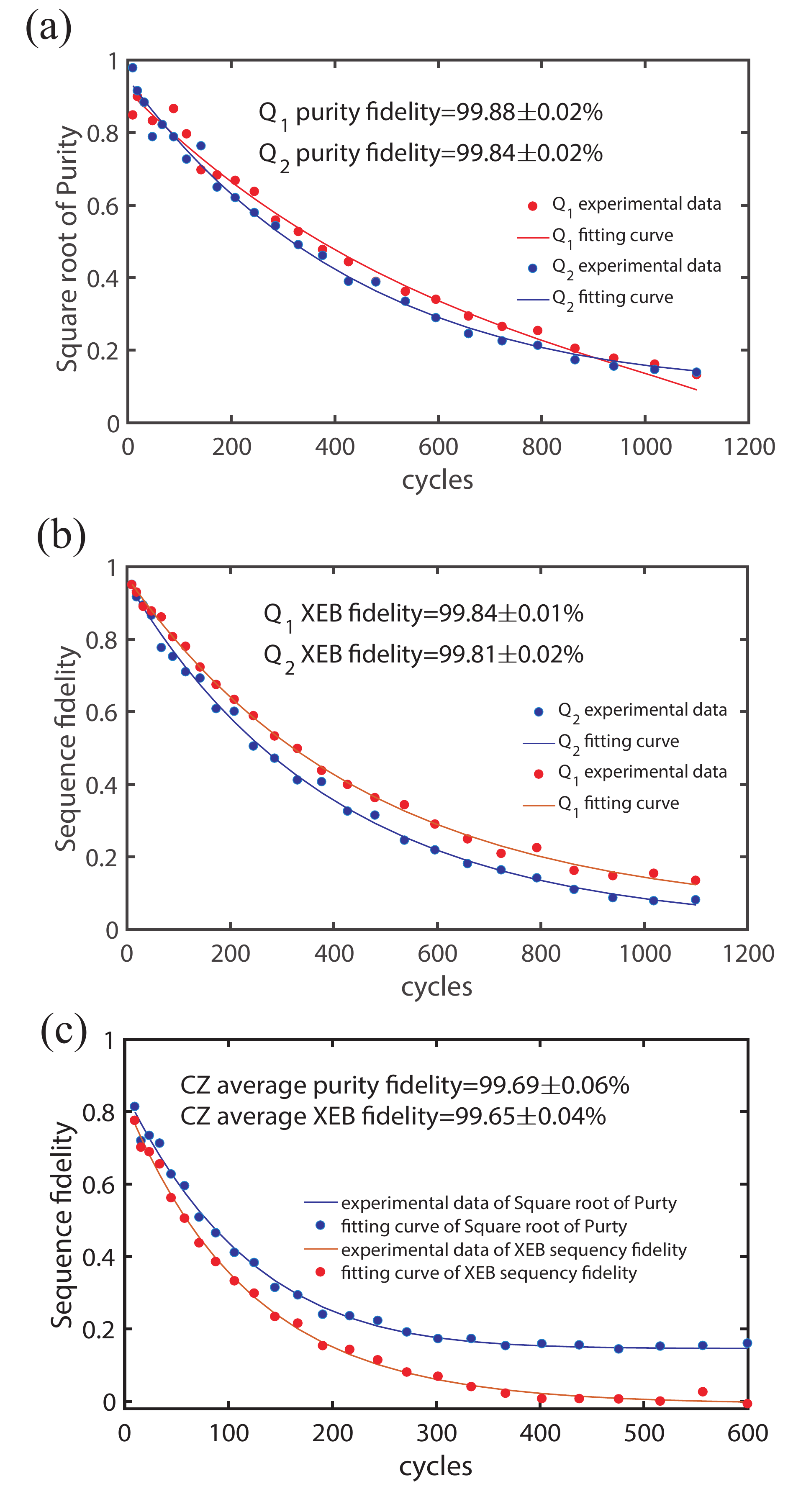}
		\caption{ (a) Pauli purity fidelities for single $\pi/2$ gates. (b)  Pauli fidelities for single qubit $\pi/2$ gates. (c) Average purity fidelity and averge fidelity of the CZ gate.}
		\label{all_xeb}
	\end{figure}

	After preparation experiments finished above,  speckle purity benchmarking (SPB) and cross entrophy benchmarking (XEB) are measured to assess the fidelity of single qubit gate and the CZ gate. The Pauli fidelities of single $\pi/2$ gates are 99.84$\pm$0.01\% and 99.81$\pm$0.02\%  and the Pauli purity fidelities of single qubit $\pi/2$ gates are 99.88$\pm$0.02\% and 99.84$\pm$0.02\% for $\text{Q}_{1}$ and $\text{Q}_{2}$ respectively as shown in Fig.~\ref{all_xeb}(b) and Fig.~\ref{all_xeb}(a). We use following unitary to express CZ gate
	\begin{equation}      
		\label{cz}
		\left(               
		\begin{array}{cccc}  
			1 & 0 & 0 & 0\\ 
			0 & e^{i(\Delta_{+}+\Delta_{-})} & 0 & 0\\ 
			0 & 0 & e^{i(\Delta_{+}-\Delta_{-})} & 0\\
			0 & 0 & 0 & e^{i(2\Delta_{+}-\pi)} \\
		\end{array}
		\right)                 
	\end{equation}
	The additional z ratations can be corrected by applying virtual Z gates that do not influence the gate fidelity~\cite{McKay}. So we use NM algorithm to fitting the rotation angles $\Delta_{+}$ and $\Delta_{-}$ and then get the average CZ gate fidelity is 99.65$\pm$0.04\% and the average purity fidelity is 99.69$\pm$0.04\% in Fig.~\ref{all_xeb}(c) which means the control error is about 0.04$\%$. We think the zpulse distortion of coupler Z bias control lines is the main reason for the control error.

	In conclusion, our work provides a path towards buliding extensible superconduting qubit system in which each qubits can tunably couple to four near-neighbor qubits. We realize high-fidelity CZ gate in this prototype system  which can improve the accuracy of various quantum simulations and promote the development of fault-tolerant quantum computation. We also raise a question that how to correct zpulse distortion of the coupler zpulse control line without corresponding readout resonators. Taken together, the designs and demonstrations will help resovle many chanllenges in the implementation of large scale quantum systems.

\bibliography{references}

\begin{thebibliography}{23}%
\makeatletter
\providecommand \@ifxundefined [1]{%
 \@ifx{#1\undefined}
}%
\providecommand \@ifnum [1]{%
 \ifnum #1\expandafter \@firstoftwo
 \else \expandafter \@secondoftwo
 \fi
}%
\providecommand \@ifx [1]{%
 \ifx #1\expandafter \@firstoftwo
 \else \expandafter \@secondoftwo
 \fi
}%
\providecommand \natexlab [1]{#1}%
\providecommand \enquote  [1]{``#1''}%
\providecommand \bibnamefont  [1]{#1}%
\providecommand \bibfnamefont [1]{#1}%
\providecommand \citenamefont [1]{#1}%
\providecommand \href@noop [0]{\@secondoftwo}%
\providecommand \href [0]{\begingroup \@sanitize@url \@href}%
\providecommand \@href[1]{\@@startlink{#1}\@@href}%
\providecommand \@@href[1]{\endgroup#1\@@endlink}%
\providecommand \@sanitize@url [0]{\catcode `\\12\catcode `\$12\catcode
  `\&12\catcode `\#12\catcode `\^12\catcode `\_12\catcode `\%12\relax}%
\providecommand \@@startlink[1]{}%
\providecommand \@@endlink[0]{}%
\providecommand \url  [0]{\begingroup\@sanitize@url \@url }%
\providecommand \@url [1]{\endgroup\@href {#1}{\urlprefix }}%
\providecommand \urlprefix  [0]{URL }%
\providecommand \Eprint [0]{\href }%
\providecommand \doibase [0]{http://dx.doi.org/}%
\providecommand \selectlanguage [0]{\@gobble}%
\providecommand \bibinfo  [0]{\@secondoftwo}%
\providecommand \bibfield  [0]{\@secondoftwo}%
\providecommand \translation [1]{[#1]}%
\providecommand \BibitemOpen [0]{}%
\providecommand \bibitemStop [0]{}%
\providecommand \bibitemNoStop [0]{.\EOS\space}%
\providecommand \EOS [0]{\spacefactor3000\relax}%
\providecommand \BibitemShut  [1]{\csname bibitem#1\endcsname}%
\let\auto@bib@innerbib\@empty
\bibitem [{\citenamefont {Krantz}\ \emph {et~al.}(2019)\citenamefont {Krantz},
  \citenamefont {Kjaergaard}, \citenamefont {Yan}, \citenamefont {Orlando},
  \citenamefont {Gustavsson},\ and\ \citenamefont {Oliver}}]{Krantz}%
  \BibitemOpen
  \bibfield  {author} {\bibinfo {author} {\bibfnamefont {P.}~\bibnamefont
  {Krantz}}, \bibinfo {author} {\bibfnamefont {M.}~\bibnamefont {Kjaergaard}},
  \bibinfo {author} {\bibfnamefont {F.}~\bibnamefont {Yan}}, \bibinfo {author}
  {\bibfnamefont {T.~P.}\ \bibnamefont {Orlando}}, \bibinfo {author}
  {\bibfnamefont {S.}~\bibnamefont {Gustavsson}}, \ and\ \bibinfo {author}
  {\bibfnamefont {W.~D.}\ \bibnamefont {Oliver}},\ }\href {\doibase
  10.1063/1.5089550} {\bibfield  {journal} {\bibinfo  {journal} {Applied
  Physics Reviews}\ }\textbf {\bibinfo {volume} {6}},\ \bibinfo {pages}
  {021318} (\bibinfo {year} {2019})}\BibitemShut {NoStop}%
\bibitem [{\citenamefont {Buluta}\ and\ \citenamefont {Nori}(2009)}]{Buluta}%
  \BibitemOpen
  \bibfield  {author} {\bibinfo {author} {\bibfnamefont {I.}~\bibnamefont
  {Buluta}}\ and\ \bibinfo {author} {\bibfnamefont {F.}~\bibnamefont {Nori}},\
  }\href {\doibase 10.1126/science.1177838} {\bibfield  {journal} {\bibinfo
  {journal} {Science}\ }\textbf {\bibinfo {volume} {326}},\ \bibinfo {pages}
  {108} (\bibinfo {year} {2009})}\BibitemShut {NoStop}%
\bibitem [{\citenamefont {Houck}\ \emph {et~al.}(2012)\citenamefont {Houck},
  \citenamefont {Türeci},\ and\ \citenamefont {Koch}}]{Houck}%
  \BibitemOpen
  \bibfield  {author} {\bibinfo {author} {\bibfnamefont {A.~A.}\ \bibnamefont
  {Houck}}, \bibinfo {author} {\bibfnamefont {H.}~\bibnamefont {Türeci}}, \
  and\ \bibinfo {author} {\bibfnamefont {J.}~\bibnamefont {Koch}},\ }\href
  {https://doi.org/10.1038/nphys2251} {\bibfield  {journal} {\bibinfo
  {journal} {Nature Physics}\ }\textbf {\bibinfo {volume} {8}},\ \bibinfo
  {pages} {292} (\bibinfo {year} {2012})}\BibitemShut {NoStop}%
\bibitem [{\citenamefont {Georgescu}\ \emph {et~al.}(2014)\citenamefont
  {Georgescu}, \citenamefont {Ashhab},\ and\ \citenamefont {Nori}}]{Georgescu}%
  \BibitemOpen
  \bibfield  {author} {\bibinfo {author} {\bibfnamefont {I.~M.}\ \bibnamefont
  {Georgescu}}, \bibinfo {author} {\bibfnamefont {S.}~\bibnamefont {Ashhab}}, \
  and\ \bibinfo {author} {\bibfnamefont {F.}~\bibnamefont {Nori}},\ }\href
  {\doibase 10.1103/RevModPhys.86.153} {\bibfield  {journal} {\bibinfo
  {journal} {Rev. Mod. Phys.}\ }\textbf {\bibinfo {volume} {86}},\ \bibinfo
  {pages} {153} (\bibinfo {year} {2014})}\BibitemShut {NoStop}%
\bibitem [{\citenamefont {Devoret}\ and\ \citenamefont
  {Schoelkopf}(2013)}]{Devoret}%
  \BibitemOpen
  \bibfield  {author} {\bibinfo {author} {\bibfnamefont {M.~H.}\ \bibnamefont
  {Devoret}}\ and\ \bibinfo {author} {\bibfnamefont {R.~J.}\ \bibnamefont
  {Schoelkopf}},\ }\href {\doibase 10.1126/science.1231930} {\bibfield
  {journal} {\bibinfo  {journal} {Science}\ }\textbf {\bibinfo {volume}
  {339}},\ \bibinfo {pages} {1169} (\bibinfo {year} {2013})}\BibitemShut
  {NoStop}%
\bibitem [{\citenamefont {Campbell}\ \emph {et~al.}(2017)\citenamefont
  {Campbell}, \citenamefont {Terhal},\ and\ \citenamefont
  {Vuillot}}]{Campbell}%
  \BibitemOpen
  \bibfield  {author} {\bibinfo {author} {\bibfnamefont {E.~T.}\ \bibnamefont
  {Campbell}}, \bibinfo {author} {\bibfnamefont {B.~M.}\ \bibnamefont
  {Terhal}}, \ and\ \bibinfo {author} {\bibfnamefont {C.}~\bibnamefont
  {Vuillot}},\ }\href {https://doi.org/10.1038/nature23460} {\bibfield
  {journal} {\bibinfo  {journal} {Nature}\ }\textbf {\bibinfo {volume} {549}},\
  \bibinfo {pages} {172} (\bibinfo {year} {2017})}\BibitemShut {NoStop}%
\bibitem [{\citenamefont {Koch}\ \emph {et~al.}(2007)\citenamefont {Koch},
  \citenamefont {Yu}, \citenamefont {Gambetta}, \citenamefont {Houck},
  \citenamefont {Schuster}, \citenamefont {Majer}, \citenamefont {Blais},
  \citenamefont {Devoret}, \citenamefont {Girvin},\ and\ \citenamefont
  {Schoelkopf}}]{Koch2007}%
  \BibitemOpen
  \bibfield  {author} {\bibinfo {author} {\bibfnamefont {J.}~\bibnamefont
  {Koch}}, \bibinfo {author} {\bibfnamefont {T.~M.}\ \bibnamefont {Yu}},
  \bibinfo {author} {\bibfnamefont {J.}~\bibnamefont {Gambetta}}, \bibinfo
  {author} {\bibfnamefont {A.~A.}\ \bibnamefont {Houck}}, \bibinfo {author}
  {\bibfnamefont {D.~I.}\ \bibnamefont {Schuster}}, \bibinfo {author}
  {\bibfnamefont {J.}~\bibnamefont {Majer}}, \bibinfo {author} {\bibfnamefont
  {A.}~\bibnamefont {Blais}}, \bibinfo {author} {\bibfnamefont {M.~H.}\
  \bibnamefont {Devoret}}, \bibinfo {author} {\bibfnamefont {S.~M.}\
  \bibnamefont {Girvin}}, \ and\ \bibinfo {author} {\bibfnamefont {R.~J.}\
  \bibnamefont {Schoelkopf}},\ }\href {\doibase 10.1103/PhysRevA.76.042319}
  {\bibfield  {journal} {\bibinfo  {journal} {Phys. Rev. A}\ }\textbf {\bibinfo
  {volume} {76}},\ \bibinfo {pages} {042319} (\bibinfo {year}
  {2007})}\BibitemShut {NoStop}%
\bibitem [{\citenamefont {Barends}\ \emph {et~al.}(2013)\citenamefont
  {Barends}, \citenamefont {Kelly}, \citenamefont {Megrant}, \citenamefont
  {Sank}, \citenamefont {Jeffrey}, \citenamefont {Chen}, \citenamefont {Yin},
  \citenamefont {Chiaro}, \citenamefont {Mutus}, \citenamefont {Neill},
  \citenamefont {O'Malley}, \citenamefont {Roushan}, \citenamefont {Wenner},
  \citenamefont {White}, \citenamefont {Cleland},\ and\ \citenamefont
  {Martinis}}]{Barends}%
  \BibitemOpen
  \bibfield  {author} {\bibinfo {author} {\bibfnamefont {R.}~\bibnamefont
  {Barends}}, \bibinfo {author} {\bibfnamefont {J.}~\bibnamefont {Kelly}},
  \bibinfo {author} {\bibfnamefont {A.}~\bibnamefont {Megrant}}, \bibinfo
  {author} {\bibfnamefont {D.}~\bibnamefont {Sank}}, \bibinfo {author}
  {\bibfnamefont {E.}~\bibnamefont {Jeffrey}}, \bibinfo {author} {\bibfnamefont
  {Y.}~\bibnamefont {Chen}}, \bibinfo {author} {\bibfnamefont {Y.}~\bibnamefont
  {Yin}}, \bibinfo {author} {\bibfnamefont {B.}~\bibnamefont {Chiaro}},
  \bibinfo {author} {\bibfnamefont {J.}~\bibnamefont {Mutus}}, \bibinfo
  {author} {\bibfnamefont {C.}~\bibnamefont {Neill}}, \bibinfo {author}
  {\bibfnamefont {P.}~\bibnamefont {O'Malley}}, \bibinfo {author}
  {\bibfnamefont {P.}~\bibnamefont {Roushan}}, \bibinfo {author} {\bibfnamefont
  {J.}~\bibnamefont {Wenner}}, \bibinfo {author} {\bibfnamefont {T.~C.}\
  \bibnamefont {White}}, \bibinfo {author} {\bibfnamefont {A.~N.}\ \bibnamefont
  {Cleland}}, \ and\ \bibinfo {author} {\bibfnamefont {J.~M.}\ \bibnamefont
  {Martinis}},\ }\href {\doibase 10.1103/PhysRevLett.111.080502} {\bibfield
  {journal} {\bibinfo  {journal} {Phys. Rev. Lett.}\ }\textbf {\bibinfo
  {volume} {111}},\ \bibinfo {pages} {080502} (\bibinfo {year}
  {2013})}\BibitemShut {NoStop}%
\bibitem [{\citenamefont {Martinis}\ and\ \citenamefont
  {Geller}(2014)}]{Martinis}%
  \BibitemOpen
  \bibfield  {author} {\bibinfo {author} {\bibfnamefont {J.~M.}\ \bibnamefont
  {Martinis}}\ and\ \bibinfo {author} {\bibfnamefont {M.~R.}\ \bibnamefont
  {Geller}},\ }\href {\doibase 10.1103/PhysRevA.90.022307} {\bibfield
  {journal} {\bibinfo  {journal} {Phys. Rev. A}\ }\textbf {\bibinfo {volume}
  {90}},\ \bibinfo {pages} {022307} (\bibinfo {year} {2014})}\BibitemShut
  {NoStop}%
\bibitem [{\citenamefont {Chen}\ \emph {et~al.}(2014)\citenamefont {Chen},
  \citenamefont {Neill}, \citenamefont {Roushan}, \citenamefont {Leung},
  \citenamefont {Fang}, \citenamefont {Barends}, \citenamefont {Kelly},
  \citenamefont {Campbell}, \citenamefont {Chen}, \citenamefont {Chiaro},
  \citenamefont {Dunsworth}, \citenamefont {Jeffrey}, \citenamefont {Megrant},
  \citenamefont {Mutus}, \citenamefont {O'Malley}, \citenamefont {Quintana},
  \citenamefont {Sank}, \citenamefont {Vainsencher}, \citenamefont {Wenner},
  \citenamefont {White}, \citenamefont {Geller}, \citenamefont {Cleland},\ and\
  \citenamefont {Martinis}}]{Chen2014}%
  \BibitemOpen
  \bibfield  {author} {\bibinfo {author} {\bibfnamefont {Y.}~\bibnamefont
  {Chen}}, \bibinfo {author} {\bibfnamefont {C.}~\bibnamefont {Neill}},
  \bibinfo {author} {\bibfnamefont {P.}~\bibnamefont {Roushan}}, \bibinfo
  {author} {\bibfnamefont {N.}~\bibnamefont {Leung}}, \bibinfo {author}
  {\bibfnamefont {M.}~\bibnamefont {Fang}}, \bibinfo {author} {\bibfnamefont
  {R.}~\bibnamefont {Barends}}, \bibinfo {author} {\bibfnamefont
  {J.}~\bibnamefont {Kelly}}, \bibinfo {author} {\bibfnamefont
  {B.}~\bibnamefont {Campbell}}, \bibinfo {author} {\bibfnamefont
  {Z.}~\bibnamefont {Chen}}, \bibinfo {author} {\bibfnamefont {B.}~\bibnamefont
  {Chiaro}}, \bibinfo {author} {\bibfnamefont {A.}~\bibnamefont {Dunsworth}},
  \bibinfo {author} {\bibfnamefont {E.}~\bibnamefont {Jeffrey}}, \bibinfo
  {author} {\bibfnamefont {A.}~\bibnamefont {Megrant}}, \bibinfo {author}
  {\bibfnamefont {J.~Y.}\ \bibnamefont {Mutus}}, \bibinfo {author}
  {\bibfnamefont {P.~J.~J.}\ \bibnamefont {O'Malley}}, \bibinfo {author}
  {\bibfnamefont {C.~M.}\ \bibnamefont {Quintana}}, \bibinfo {author}
  {\bibfnamefont {D.}~\bibnamefont {Sank}}, \bibinfo {author} {\bibfnamefont
  {A.}~\bibnamefont {Vainsencher}}, \bibinfo {author} {\bibfnamefont
  {J.}~\bibnamefont {Wenner}}, \bibinfo {author} {\bibfnamefont {T.~C.}\
  \bibnamefont {White}}, \bibinfo {author} {\bibfnamefont {M.~R.}\ \bibnamefont
  {Geller}}, \bibinfo {author} {\bibfnamefont {A.~N.}\ \bibnamefont {Cleland}},
  \ and\ \bibinfo {author} {\bibfnamefont {J.~M.}\ \bibnamefont {Martinis}},\
  }\href {\doibase 10.1103/PhysRevLett.113.220502} {\bibfield  {journal}
  {\bibinfo  {journal} {Phys. Rev. Lett.}\ }\textbf {\bibinfo {volume} {113}},\
  \bibinfo {pages} {220502} (\bibinfo {year} {2014})}\BibitemShut {NoStop}%
\bibitem [{\citenamefont {Yan}\ \emph {et~al.}(2018)\citenamefont {Yan},
  \citenamefont {Krantz}, \citenamefont {Sung}, \citenamefont {Kjaergaard},
  \citenamefont {Campbell}, \citenamefont {Orlando}, \citenamefont
  {Gustavsson},\ and\ \citenamefont {Oliver}}]{YanF2018}%
  \BibitemOpen
  \bibfield  {author} {\bibinfo {author} {\bibfnamefont {F.}~\bibnamefont
  {Yan}}, \bibinfo {author} {\bibfnamefont {P.}~\bibnamefont {Krantz}},
  \bibinfo {author} {\bibfnamefont {Y.}~\bibnamefont {Sung}}, \bibinfo {author}
  {\bibfnamefont {M.}~\bibnamefont {Kjaergaard}}, \bibinfo {author}
  {\bibfnamefont {D.~L.}\ \bibnamefont {Campbell}}, \bibinfo {author}
  {\bibfnamefont {T.~P.}\ \bibnamefont {Orlando}}, \bibinfo {author}
  {\bibfnamefont {S.}~\bibnamefont {Gustavsson}}, \ and\ \bibinfo {author}
  {\bibfnamefont {W.~D.}\ \bibnamefont {Oliver}},\ }\href {\doibase
  10.1103/PhysRevApplied.10.054062} {\bibfield  {journal} {\bibinfo  {journal}
  {Phys. Rev. Applied}\ }\textbf {\bibinfo {volume} {10}},\ \bibinfo {pages}
  {054062} (\bibinfo {year} {2018})}\BibitemShut {NoStop}%
\bibitem [{\citenamefont {Niskanen}\ \emph {et~al.}(2007)\citenamefont
  {Niskanen}, \citenamefont {Harrabi}, \citenamefont {Yoshihara}, \citenamefont
  {Nakamura}, \citenamefont {Lloyd},\ and\ \citenamefont {Tsai}}]{Niskanen}%
  \BibitemOpen
  \bibfield  {author} {\bibinfo {author} {\bibfnamefont {A.~O.}\ \bibnamefont
  {Niskanen}}, \bibinfo {author} {\bibfnamefont {K.}~\bibnamefont {Harrabi}},
  \bibinfo {author} {\bibfnamefont {F.}~\bibnamefont {Yoshihara}}, \bibinfo
  {author} {\bibfnamefont {Y.}~\bibnamefont {Nakamura}}, \bibinfo {author}
  {\bibfnamefont {S.}~\bibnamefont {Lloyd}}, \ and\ \bibinfo {author}
  {\bibfnamefont {J.~S.}\ \bibnamefont {Tsai}},\ }\href {\doibase
  10.1126/science.1141324} {\bibfield  {journal} {\bibinfo  {journal}
  {Science}\ }\textbf {\bibinfo {volume} {316}},\ \bibinfo {pages} {723}
  (\bibinfo {year} {2007})}\BibitemShut {NoStop}%
\bibitem [{\citenamefont {Hime}\ \emph {et~al.}(2006)\citenamefont {Hime},
  \citenamefont {Reichardt}, \citenamefont {Plourde}, \citenamefont
  {Robertson}, \citenamefont {Wu}, \citenamefont {Ustinov},\ and\ \citenamefont
  {Clarke}}]{Hime}%
  \BibitemOpen
  \bibfield  {author} {\bibinfo {author} {\bibfnamefont {T.}~\bibnamefont
  {Hime}}, \bibinfo {author} {\bibfnamefont {P.~A.}\ \bibnamefont {Reichardt}},
  \bibinfo {author} {\bibfnamefont {B.~L.~T.}\ \bibnamefont {Plourde}},
  \bibinfo {author} {\bibfnamefont {T.~L.}\ \bibnamefont {Robertson}}, \bibinfo
  {author} {\bibfnamefont {C.-E.}\ \bibnamefont {Wu}}, \bibinfo {author}
  {\bibfnamefont {A.~V.}\ \bibnamefont {Ustinov}}, \ and\ \bibinfo {author}
  {\bibfnamefont {J.}~\bibnamefont {Clarke}},\ }\href {\doibase
  10.1126/science.1134388} {\bibfield  {journal} {\bibinfo  {journal}
  {Science}\ }\textbf {\bibinfo {volume} {314}},\ \bibinfo {pages} {1427}
  (\bibinfo {year} {2006})}\BibitemShut {NoStop}%
\bibitem [{\citenamefont {Sung}\ \emph {et~al.}(2021)\citenamefont {Sung},
  \citenamefont {Ding}, \citenamefont {Braum\"uller}, \citenamefont
  {Veps\"al\"ainen}, \citenamefont {Kannan}, \citenamefont {Kjaergaard},
  \citenamefont {Greene}, \citenamefont {Samach}, \citenamefont {McNally},
  \citenamefont {Kim}, \citenamefont {Melville}, \citenamefont {Niedzielski},
  \citenamefont {Schwartz}, \citenamefont {Yoder}, \citenamefont {Orlando},
  \citenamefont {Gustavsson},\ and\ \citenamefont {Oliver}}]{sung2020}%
  \BibitemOpen
  \bibfield  {author} {\bibinfo {author} {\bibfnamefont {Y.}~\bibnamefont
  {Sung}}, \bibinfo {author} {\bibfnamefont {L.}~\bibnamefont {Ding}}, \bibinfo
  {author} {\bibfnamefont {J.}~\bibnamefont {Braum\"uller}}, \bibinfo {author}
  {\bibfnamefont {A.}~\bibnamefont {Veps\"al\"ainen}}, \bibinfo {author}
  {\bibfnamefont {B.}~\bibnamefont {Kannan}}, \bibinfo {author} {\bibfnamefont
  {M.}~\bibnamefont {Kjaergaard}}, \bibinfo {author} {\bibfnamefont
  {A.}~\bibnamefont {Greene}}, \bibinfo {author} {\bibfnamefont {G.~O.}\
  \bibnamefont {Samach}}, \bibinfo {author} {\bibfnamefont {C.}~\bibnamefont
  {McNally}}, \bibinfo {author} {\bibfnamefont {D.}~\bibnamefont {Kim}},
  \bibinfo {author} {\bibfnamefont {A.}~\bibnamefont {Melville}}, \bibinfo
  {author} {\bibfnamefont {B.~M.}\ \bibnamefont {Niedzielski}}, \bibinfo
  {author} {\bibfnamefont {M.~E.}\ \bibnamefont {Schwartz}}, \bibinfo {author}
  {\bibfnamefont {J.~L.}\ \bibnamefont {Yoder}}, \bibinfo {author}
  {\bibfnamefont {T.~P.}\ \bibnamefont {Orlando}}, \bibinfo {author}
  {\bibfnamefont {S.}~\bibnamefont {Gustavsson}}, \ and\ \bibinfo {author}
  {\bibfnamefont {W.~D.}\ \bibnamefont {Oliver}},\ }\href {\doibase
  10.1103/PhysRevX.11.021058} {\bibfield  {journal} {\bibinfo  {journal} {Phys.
  Rev. X}\ }\textbf {\bibinfo {volume} {11}},\ \bibinfo {pages} {021058}
  (\bibinfo {year} {2021})}\BibitemShut {NoStop}%
\bibitem [{\citenamefont {Foxen}\ \emph {et~al.}(2020)\citenamefont {Foxen},
  \citenamefont {Neill}, \citenamefont {Dunsworth}, \citenamefont {Roushan},
  \citenamefont {Chiaro}, \citenamefont {Megrant}, \citenamefont {Kelly},
  \citenamefont {Chen}, \citenamefont {Satzinger}, \citenamefont {Barends},
  \citenamefont {Arute}, \citenamefont {Arya}, \citenamefont {Babbush},
  \citenamefont {Bacon}, \citenamefont {Bardin}, \citenamefont {Boixo},
  \citenamefont {Buell}, \citenamefont {Burkett}, \citenamefont {Chen},
  \citenamefont {Collins}, \citenamefont {Farhi}, \citenamefont {Fowler},
  \citenamefont {Gidney}, \citenamefont {Giustina}, \citenamefont {Graff},
  \citenamefont {Harrigan}, \citenamefont {Huang}, \citenamefont {Isakov},
  \citenamefont {Jeffrey}, \citenamefont {Jiang}, \citenamefont {Kafri},
  \citenamefont {Kechedzhi}, \citenamefont {Klimov}, \citenamefont {Korotkov},
  \citenamefont {Kostritsa}, \citenamefont {Landhuis}, \citenamefont {Lucero},
  \citenamefont {McClean}, \citenamefont {McEwen}, \citenamefont {Mi},
  \citenamefont {Mohseni}, \citenamefont {Mutus}, \citenamefont {Naaman},
  \citenamefont {Neeley}, \citenamefont {Niu}, \citenamefont {Petukhov},
  \citenamefont {Quintana}, \citenamefont {Rubin}, \citenamefont {Sank},
  \citenamefont {Smelyanskiy}, \citenamefont {Vainsencher}, \citenamefont
  {White}, \citenamefont {Yao}, \citenamefont {Yeh}, \citenamefont {Zalcman},
  \citenamefont {Neven},\ and\ \citenamefont {Martinis}}]{Foxen}%
  \BibitemOpen
  \bibfield  {author} {\bibinfo {author} {\bibfnamefont {B.}~\bibnamefont
  {Foxen}}, \bibinfo {author} {\bibfnamefont {C.}~\bibnamefont {Neill}},
  \bibinfo {author} {\bibfnamefont {A.}~\bibnamefont {Dunsworth}}, \bibinfo
  {author} {\bibfnamefont {P.}~\bibnamefont {Roushan}}, \bibinfo {author}
  {\bibfnamefont {B.}~\bibnamefont {Chiaro}}, \bibinfo {author} {\bibfnamefont
  {A.}~\bibnamefont {Megrant}}, \bibinfo {author} {\bibfnamefont
  {J.}~\bibnamefont {Kelly}}, \bibinfo {author} {\bibfnamefont
  {Z.}~\bibnamefont {Chen}}, \bibinfo {author} {\bibfnamefont {K.}~\bibnamefont
  {Satzinger}}, \bibinfo {author} {\bibfnamefont {R.}~\bibnamefont {Barends}},
  \bibinfo {author} {\bibfnamefont {F.}~\bibnamefont {Arute}}, \bibinfo
  {author} {\bibfnamefont {K.}~\bibnamefont {Arya}}, \bibinfo {author}
  {\bibfnamefont {R.}~\bibnamefont {Babbush}}, \bibinfo {author} {\bibfnamefont
  {D.}~\bibnamefont {Bacon}}, \bibinfo {author} {\bibfnamefont {J.~C.}\
  \bibnamefont {Bardin}}, \bibinfo {author} {\bibfnamefont {S.}~\bibnamefont
  {Boixo}}, \bibinfo {author} {\bibfnamefont {D.}~\bibnamefont {Buell}},
  \bibinfo {author} {\bibfnamefont {B.}~\bibnamefont {Burkett}}, \bibinfo
  {author} {\bibfnamefont {Y.}~\bibnamefont {Chen}}, \bibinfo {author}
  {\bibfnamefont {R.}~\bibnamefont {Collins}}, \bibinfo {author} {\bibfnamefont
  {E.}~\bibnamefont {Farhi}}, \bibinfo {author} {\bibfnamefont
  {A.}~\bibnamefont {Fowler}}, \bibinfo {author} {\bibfnamefont
  {C.}~\bibnamefont {Gidney}}, \bibinfo {author} {\bibfnamefont
  {M.}~\bibnamefont {Giustina}}, \bibinfo {author} {\bibfnamefont
  {R.}~\bibnamefont {Graff}}, \bibinfo {author} {\bibfnamefont
  {M.}~\bibnamefont {Harrigan}}, \bibinfo {author} {\bibfnamefont
  {T.}~\bibnamefont {Huang}}, \bibinfo {author} {\bibfnamefont {S.~V.}\
  \bibnamefont {Isakov}}, \bibinfo {author} {\bibfnamefont {E.}~\bibnamefont
  {Jeffrey}}, \bibinfo {author} {\bibfnamefont {Z.}~\bibnamefont {Jiang}},
  \bibinfo {author} {\bibfnamefont {D.}~\bibnamefont {Kafri}}, \bibinfo
  {author} {\bibfnamefont {K.}~\bibnamefont {Kechedzhi}}, \bibinfo {author}
  {\bibfnamefont {P.}~\bibnamefont {Klimov}}, \bibinfo {author} {\bibfnamefont
  {A.}~\bibnamefont {Korotkov}}, \bibinfo {author} {\bibfnamefont
  {F.}~\bibnamefont {Kostritsa}}, \bibinfo {author} {\bibfnamefont
  {D.}~\bibnamefont {Landhuis}}, \bibinfo {author} {\bibfnamefont
  {E.}~\bibnamefont {Lucero}}, \bibinfo {author} {\bibfnamefont
  {J.}~\bibnamefont {McClean}}, \bibinfo {author} {\bibfnamefont
  {M.}~\bibnamefont {McEwen}}, \bibinfo {author} {\bibfnamefont
  {X.}~\bibnamefont {Mi}}, \bibinfo {author} {\bibfnamefont {M.}~\bibnamefont
  {Mohseni}}, \bibinfo {author} {\bibfnamefont {J.~Y.}\ \bibnamefont {Mutus}},
  \bibinfo {author} {\bibfnamefont {O.}~\bibnamefont {Naaman}}, \bibinfo
  {author} {\bibfnamefont {M.}~\bibnamefont {Neeley}}, \bibinfo {author}
  {\bibfnamefont {M.}~\bibnamefont {Niu}}, \bibinfo {author} {\bibfnamefont
  {A.}~\bibnamefont {Petukhov}}, \bibinfo {author} {\bibfnamefont
  {C.}~\bibnamefont {Quintana}}, \bibinfo {author} {\bibfnamefont
  {N.}~\bibnamefont {Rubin}}, \bibinfo {author} {\bibfnamefont
  {D.}~\bibnamefont {Sank}}, \bibinfo {author} {\bibfnamefont {V.}~\bibnamefont
  {Smelyanskiy}}, \bibinfo {author} {\bibfnamefont {A.}~\bibnamefont
  {Vainsencher}}, \bibinfo {author} {\bibfnamefont {T.~C.}\ \bibnamefont
  {White}}, \bibinfo {author} {\bibfnamefont {Z.}~\bibnamefont {Yao}}, \bibinfo
  {author} {\bibfnamefont {P.}~\bibnamefont {Yeh}}, \bibinfo {author}
  {\bibfnamefont {A.}~\bibnamefont {Zalcman}}, \bibinfo {author} {\bibfnamefont
  {H.}~\bibnamefont {Neven}}, \ and\ \bibinfo {author} {\bibfnamefont {J.~M.}\
  \bibnamefont {Martinis}} (\bibinfo {collaboration} {Google AI Quantum}),\
  }\href {\doibase 10.1103/PhysRevLett.125.120504} {\bibfield  {journal}
  {\bibinfo  {journal} {Phys. Rev. Lett.}\ }\textbf {\bibinfo {volume} {125}},\
  \bibinfo {pages} {120504} (\bibinfo {year} {2020})}\BibitemShut {NoStop}%
\bibitem [{\citenamefont {Barends}\ \emph {et~al.}(2019)\citenamefont
  {Barends}, \citenamefont {Quintana}, \citenamefont {Petukhov}, \citenamefont
  {Chen}, \citenamefont {Kafri}, \citenamefont {Kechedzhi}, \citenamefont
  {Collins}, \citenamefont {Naaman}, \citenamefont {Boixo}, \citenamefont
  {Arute}, \citenamefont {Arya}, \citenamefont {Buell}, \citenamefont
  {Burkett}, \citenamefont {Chen}, \citenamefont {Chiaro}, \citenamefont
  {Dunsworth}, \citenamefont {Foxen}, \citenamefont {Fowler}, \citenamefont
  {Gidney}, \citenamefont {Giustina}, \citenamefont {Graff}, \citenamefont
  {Huang}, \citenamefont {Jeffrey}, \citenamefont {Kelly}, \citenamefont
  {Klimov}, \citenamefont {Kostritsa}, \citenamefont {Landhuis}, \citenamefont
  {Lucero}, \citenamefont {McEwen}, \citenamefont {Megrant}, \citenamefont
  {Mi}, \citenamefont {Mutus}, \citenamefont {Neeley}, \citenamefont {Neill},
  \citenamefont {Ostby}, \citenamefont {Roushan}, \citenamefont {Sank},
  \citenamefont {Satzinger}, \citenamefont {Vainsencher}, \citenamefont
  {White}, \citenamefont {Yao}, \citenamefont {Yeh}, \citenamefont {Zalcman},
  \citenamefont {Neven}, \citenamefont {Smelyanskiy},\ and\ \citenamefont
  {Martinis}}]{DgateXEB2019}%
  \BibitemOpen
  \bibfield  {author} {\bibinfo {author} {\bibfnamefont {R.}~\bibnamefont
  {Barends}}, \bibinfo {author} {\bibfnamefont {C.~M.}\ \bibnamefont
  {Quintana}}, \bibinfo {author} {\bibfnamefont {A.~G.}\ \bibnamefont
  {Petukhov}}, \bibinfo {author} {\bibfnamefont {Y.}~\bibnamefont {Chen}},
  \bibinfo {author} {\bibfnamefont {D.}~\bibnamefont {Kafri}}, \bibinfo
  {author} {\bibfnamefont {K.}~\bibnamefont {Kechedzhi}}, \bibinfo {author}
  {\bibfnamefont {R.}~\bibnamefont {Collins}}, \bibinfo {author} {\bibfnamefont
  {O.}~\bibnamefont {Naaman}}, \bibinfo {author} {\bibfnamefont
  {S.}~\bibnamefont {Boixo}}, \bibinfo {author} {\bibfnamefont
  {F.}~\bibnamefont {Arute}}, \bibinfo {author} {\bibfnamefont
  {K.}~\bibnamefont {Arya}}, \bibinfo {author} {\bibfnamefont {D.}~\bibnamefont
  {Buell}}, \bibinfo {author} {\bibfnamefont {B.}~\bibnamefont {Burkett}},
  \bibinfo {author} {\bibfnamefont {Z.}~\bibnamefont {Chen}}, \bibinfo {author}
  {\bibfnamefont {B.}~\bibnamefont {Chiaro}}, \bibinfo {author} {\bibfnamefont
  {A.}~\bibnamefont {Dunsworth}}, \bibinfo {author} {\bibfnamefont
  {B.}~\bibnamefont {Foxen}}, \bibinfo {author} {\bibfnamefont
  {A.}~\bibnamefont {Fowler}}, \bibinfo {author} {\bibfnamefont
  {C.}~\bibnamefont {Gidney}}, \bibinfo {author} {\bibfnamefont
  {M.}~\bibnamefont {Giustina}}, \bibinfo {author} {\bibfnamefont
  {R.}~\bibnamefont {Graff}}, \bibinfo {author} {\bibfnamefont
  {T.}~\bibnamefont {Huang}}, \bibinfo {author} {\bibfnamefont
  {E.}~\bibnamefont {Jeffrey}}, \bibinfo {author} {\bibfnamefont
  {J.}~\bibnamefont {Kelly}}, \bibinfo {author} {\bibfnamefont {P.~V.}\
  \bibnamefont {Klimov}}, \bibinfo {author} {\bibfnamefont {F.}~\bibnamefont
  {Kostritsa}}, \bibinfo {author} {\bibfnamefont {D.}~\bibnamefont {Landhuis}},
  \bibinfo {author} {\bibfnamefont {E.}~\bibnamefont {Lucero}}, \bibinfo
  {author} {\bibfnamefont {M.}~\bibnamefont {McEwen}}, \bibinfo {author}
  {\bibfnamefont {A.}~\bibnamefont {Megrant}}, \bibinfo {author} {\bibfnamefont
  {X.}~\bibnamefont {Mi}}, \bibinfo {author} {\bibfnamefont {J.}~\bibnamefont
  {Mutus}}, \bibinfo {author} {\bibfnamefont {M.}~\bibnamefont {Neeley}},
  \bibinfo {author} {\bibfnamefont {C.}~\bibnamefont {Neill}}, \bibinfo
  {author} {\bibfnamefont {E.}~\bibnamefont {Ostby}}, \bibinfo {author}
  {\bibfnamefont {P.}~\bibnamefont {Roushan}}, \bibinfo {author} {\bibfnamefont
  {D.}~\bibnamefont {Sank}}, \bibinfo {author} {\bibfnamefont {K.~J.}\
  \bibnamefont {Satzinger}}, \bibinfo {author} {\bibfnamefont {A.}~\bibnamefont
  {Vainsencher}}, \bibinfo {author} {\bibfnamefont {T.}~\bibnamefont {White}},
  \bibinfo {author} {\bibfnamefont {J.}~\bibnamefont {Yao}}, \bibinfo {author}
  {\bibfnamefont {P.}~\bibnamefont {Yeh}}, \bibinfo {author} {\bibfnamefont
  {A.}~\bibnamefont {Zalcman}}, \bibinfo {author} {\bibfnamefont
  {H.}~\bibnamefont {Neven}}, \bibinfo {author} {\bibfnamefont {V.~N.}\
  \bibnamefont {Smelyanskiy}}, \ and\ \bibinfo {author} {\bibfnamefont {J.~M.}\
  \bibnamefont {Martinis}},\ }\href {\doibase 10.1103/PhysRevLett.123.210501}
  {\bibfield  {journal} {\bibinfo  {journal} {Phys. Rev. Lett.}\ }\textbf
  {\bibinfo {volume} {123}},\ \bibinfo {pages} {210501} (\bibinfo {year}
  {2019})}\BibitemShut {NoStop}%
\bibitem [{\citenamefont {Boixo}\ \emph {et~al.}(2018)\citenamefont {Boixo},
  \citenamefont {Isakov}, \citenamefont {Smelyanskiy}, \citenamefont {Babbush},
  \citenamefont {Ding}, \citenamefont {Jiang}, \citenamefont {Bremner},
  \citenamefont {Martinis},\ and\ \citenamefont {Neven}}]{XEB2018}%
  \BibitemOpen
  \bibfield  {author} {\bibinfo {author} {\bibfnamefont {S.}~\bibnamefont
  {Boixo}}, \bibinfo {author} {\bibfnamefont {S.~V.}\ \bibnamefont {Isakov}},
  \bibinfo {author} {\bibfnamefont {V.~N.}\ \bibnamefont {Smelyanskiy}},
  \bibinfo {author} {\bibfnamefont {R.}~\bibnamefont {Babbush}}, \bibinfo
  {author} {\bibfnamefont {N.}~\bibnamefont {Ding}}, \bibinfo {author}
  {\bibfnamefont {Z.}~\bibnamefont {Jiang}}, \bibinfo {author} {\bibfnamefont
  {M.~J.}\ \bibnamefont {Bremner}}, \bibinfo {author} {\bibfnamefont {J.~M.}\
  \bibnamefont {Martinis}}, \ and\ \bibinfo {author} {\bibfnamefont
  {H.}~\bibnamefont {Neven}},\ }\href {\doibase 10.1038/s41567-018-0124-x}
  {\bibfield  {journal} {\bibinfo  {journal} {Nature Physics}\ }\textbf
  {\bibinfo {volume} {14}},\ \bibinfo {pages} {595} (\bibinfo {year}
  {2018})}\BibitemShut {NoStop}%
\bibitem [{\citenamefont {Arute}\ \emph {et~al.}(2019)\citenamefont {Arute},
  \citenamefont {Arya}, \citenamefont {Babbush}, \citenamefont {Bacon},
  \citenamefont {Bardin}, \citenamefont {Barends}, \citenamefont {Biswas},
  \citenamefont {Boixo}, \citenamefont {Brandao}, \citenamefont {Buell},
  \citenamefont {Burkett}, \citenamefont {Chen}, \citenamefont {Chen},
  \citenamefont {Chiaro}, \citenamefont {Collins}, \citenamefont {Courtney},
  \citenamefont {Dunsworth}, \citenamefont {Farhi}, \citenamefont {Foxen},\
  and\ \citenamefont {Martinis}}]{Google-supermacy2019}%
  \BibitemOpen
  \bibfield  {author} {\bibinfo {author} {\bibfnamefont {F.}~\bibnamefont
  {Arute}}, \bibinfo {author} {\bibfnamefont {K.}~\bibnamefont {Arya}},
  \bibinfo {author} {\bibfnamefont {R.}~\bibnamefont {Babbush}}, \bibinfo
  {author} {\bibfnamefont {D.}~\bibnamefont {Bacon}}, \bibinfo {author}
  {\bibfnamefont {J.}~\bibnamefont {Bardin}}, \bibinfo {author} {\bibfnamefont
  {R.}~\bibnamefont {Barends}}, \bibinfo {author} {\bibfnamefont
  {R.}~\bibnamefont {Biswas}}, \bibinfo {author} {\bibfnamefont
  {S.}~\bibnamefont {Boixo}}, \bibinfo {author} {\bibfnamefont
  {F.}~\bibnamefont {Brandao}}, \bibinfo {author} {\bibfnamefont
  {D.}~\bibnamefont {Buell}}, \bibinfo {author} {\bibfnamefont
  {B.}~\bibnamefont {Burkett}}, \bibinfo {author} {\bibfnamefont
  {Y.}~\bibnamefont {Chen}}, \bibinfo {author} {\bibfnamefont {Z.}~\bibnamefont
  {Chen}}, \bibinfo {author} {\bibfnamefont {B.}~\bibnamefont {Chiaro}},
  \bibinfo {author} {\bibfnamefont {R.}~\bibnamefont {Collins}}, \bibinfo
  {author} {\bibfnamefont {W.}~\bibnamefont {Courtney}}, \bibinfo {author}
  {\bibfnamefont {A.}~\bibnamefont {Dunsworth}}, \bibinfo {author}
  {\bibfnamefont {E.}~\bibnamefont {Farhi}}, \bibinfo {author} {\bibfnamefont
  {B.}~\bibnamefont {Foxen}}, \ and\ \bibinfo {author} {\bibfnamefont
  {J.}~\bibnamefont {Martinis}},\ }\href {\doibase 10.1038/s41586-019-1666-5}
  {\bibfield  {journal} {\bibinfo  {journal} {Nature}\ }\textbf {\bibinfo
  {volume} {574}},\ \bibinfo {pages} {505} (\bibinfo {year}
  {2019})}\BibitemShut {NoStop}%
\bibitem [{\citenamefont {Mutus}\ \emph {et~al.}(2014)\citenamefont {Mutus},
  \citenamefont {White}, \citenamefont {Barends}, \citenamefont {Chen},
  \citenamefont {Chen}, \citenamefont {Chiaro}, \citenamefont {Dunsworth},
  \citenamefont {Jeffrey}, \citenamefont {Kelly}, \citenamefont {Megrant},
  \citenamefont {Neill}, \citenamefont {O'Malley}, \citenamefont {Roushan},
  \citenamefont {Sank}, \citenamefont {Vainsencher}, \citenamefont {Wenner},
  \citenamefont {Sundqvist}, \citenamefont {Cleland},\ and\ \citenamefont
  {Martinis}}]{JPA2014}%
  \BibitemOpen
  \bibfield  {author} {\bibinfo {author} {\bibfnamefont {J.~Y.}\ \bibnamefont
  {Mutus}}, \bibinfo {author} {\bibfnamefont {T.~C.}\ \bibnamefont {White}},
  \bibinfo {author} {\bibfnamefont {R.}~\bibnamefont {Barends}}, \bibinfo
  {author} {\bibfnamefont {Y.}~\bibnamefont {Chen}}, \bibinfo {author}
  {\bibfnamefont {Z.}~\bibnamefont {Chen}}, \bibinfo {author} {\bibfnamefont
  {B.}~\bibnamefont {Chiaro}}, \bibinfo {author} {\bibfnamefont
  {A.}~\bibnamefont {Dunsworth}}, \bibinfo {author} {\bibfnamefont
  {E.}~\bibnamefont {Jeffrey}}, \bibinfo {author} {\bibfnamefont
  {J.}~\bibnamefont {Kelly}}, \bibinfo {author} {\bibfnamefont
  {A.}~\bibnamefont {Megrant}}, \bibinfo {author} {\bibfnamefont
  {C.}~\bibnamefont {Neill}}, \bibinfo {author} {\bibfnamefont {P.~J.~J.}\
  \bibnamefont {O'Malley}}, \bibinfo {author} {\bibfnamefont {P.}~\bibnamefont
  {Roushan}}, \bibinfo {author} {\bibfnamefont {D.}~\bibnamefont {Sank}},
  \bibinfo {author} {\bibfnamefont {A.}~\bibnamefont {Vainsencher}}, \bibinfo
  {author} {\bibfnamefont {J.}~\bibnamefont {Wenner}}, \bibinfo {author}
  {\bibfnamefont {K.~M.}\ \bibnamefont {Sundqvist}}, \bibinfo {author}
  {\bibfnamefont {A.~N.}\ \bibnamefont {Cleland}}, \ and\ \bibinfo {author}
  {\bibfnamefont {J.~M.}\ \bibnamefont {Martinis}},\ }\href {\doibase
  10.1063/1.4886408} {\bibfield  {journal} {\bibinfo  {journal} {Applied
  Physics Letters}\ }\textbf {\bibinfo {volume} {104}},\ \bibinfo {pages}
  {263513} (\bibinfo {year} {2014})}\BibitemShut {NoStop}%
\bibitem [{\citenamefont {Johansson}\ \emph {et~al.}(2011)\citenamefont
  {Johansson}, \citenamefont {Nation},\ and\ \citenamefont {Nori}}]{QuTiP1}%
  \BibitemOpen
  \bibfield  {author} {\bibinfo {author} {\bibfnamefont {R.}~\bibnamefont
  {Johansson}}, \bibinfo {author} {\bibfnamefont {P.}~\bibnamefont {Nation}}, \
  and\ \bibinfo {author} {\bibfnamefont {F.}~\bibnamefont {Nori}},\ }\href
  {\doibase 10.1016/j.cpc.2012.02.021} {\bibfield  {journal} {\bibinfo
  {journal} {Computer Physics Communications}\ }\textbf {\bibinfo {volume}
  {183}},\ \bibinfo {pages} {1760} (\bibinfo {year} {2011})}\BibitemShut
  {NoStop}%
\bibitem [{\citenamefont {Johansson}\ \emph {et~al.}(2013)\citenamefont
  {Johansson}, \citenamefont {Nation},\ and\ \citenamefont {Nori}}]{QuTiP2}%
  \BibitemOpen
  \bibfield  {author} {\bibinfo {author} {\bibfnamefont {R.}~\bibnamefont
  {Johansson}}, \bibinfo {author} {\bibfnamefont {P.}~\bibnamefont {Nation}}, \
  and\ \bibinfo {author} {\bibfnamefont {F.}~\bibnamefont {Nori}},\ }\href
  {\doibase 10.1016/j.cpc.2012.11.019} {\bibfield  {journal} {\bibinfo
  {journal} {Computer Physics Communications}\ }\textbf {\bibinfo {volume}
  {184}},\ \bibinfo {pages} {1234} (\bibinfo {year} {2013})}\BibitemShut
  {NoStop}%
\bibitem [{\citenamefont {Yan}\ \emph {et~al.}(2019)\citenamefont {Yan},
  \citenamefont {Zhang}, \citenamefont {Gong}, \citenamefont {Wu},
  \citenamefont {Zheng}, \citenamefont {Li}, \citenamefont {Wang},
  \citenamefont {Liang}, \citenamefont {Lin}, \citenamefont {Xu}, \citenamefont
  {Guo}, \citenamefont {Sun}, \citenamefont {Peng}, \citenamefont {Xia},
  \citenamefont {Deng}, \citenamefont {Rong}, \citenamefont {You},
  \citenamefont {Nori}, \citenamefont {Fan}, \citenamefont {Zhu},\ and\
  \citenamefont {Pan}}]{Yan}%
  \BibitemOpen
  \bibfield  {author} {\bibinfo {author} {\bibfnamefont {Z.}~\bibnamefont
  {Yan}}, \bibinfo {author} {\bibfnamefont {Y.-R.}\ \bibnamefont {Zhang}},
  \bibinfo {author} {\bibfnamefont {M.}~\bibnamefont {Gong}}, \bibinfo {author}
  {\bibfnamefont {Y.}~\bibnamefont {Wu}}, \bibinfo {author} {\bibfnamefont
  {Y.}~\bibnamefont {Zheng}}, \bibinfo {author} {\bibfnamefont
  {S.}~\bibnamefont {Li}}, \bibinfo {author} {\bibfnamefont {C.}~\bibnamefont
  {Wang}}, \bibinfo {author} {\bibfnamefont {F.}~\bibnamefont {Liang}},
  \bibinfo {author} {\bibfnamefont {J.}~\bibnamefont {Lin}}, \bibinfo {author}
  {\bibfnamefont {Y.}~\bibnamefont {Xu}}, \bibinfo {author} {\bibfnamefont
  {C.}~\bibnamefont {Guo}}, \bibinfo {author} {\bibfnamefont {L.}~\bibnamefont
  {Sun}}, \bibinfo {author} {\bibfnamefont {C.-Z.}\ \bibnamefont {Peng}},
  \bibinfo {author} {\bibfnamefont {K.}~\bibnamefont {Xia}}, \bibinfo {author}
  {\bibfnamefont {H.}~\bibnamefont {Deng}}, \bibinfo {author} {\bibfnamefont
  {H.}~\bibnamefont {Rong}}, \bibinfo {author} {\bibfnamefont {J.~Q.}\
  \bibnamefont {You}}, \bibinfo {author} {\bibfnamefont {F.}~\bibnamefont
  {Nori}}, \bibinfo {author} {\bibfnamefont {H.}~\bibnamefont {Fan}}, \bibinfo
  {author} {\bibfnamefont {X.}~\bibnamefont {Zhu}}, \ and\ \bibinfo {author}
  {\bibfnamefont {J.-W.}\ \bibnamefont {Pan}},\ }\href {\doibase
  10.1126/science.aaw1611} {\bibfield  {journal} {\bibinfo  {journal}
  {Science}\ }\textbf {\bibinfo {volume} {364}},\ \bibinfo {pages} {753}
  (\bibinfo {year} {2019})}\BibitemShut {NoStop}%
\bibitem [{\citenamefont {McKay}\ \emph {et~al.}(2017)\citenamefont {McKay},
  \citenamefont {Wood}, \citenamefont {Sheldon}, \citenamefont {Chow},\ and\
  \citenamefont {Gambetta}}]{McKay}%
  \BibitemOpen
  \bibfield  {author} {\bibinfo {author} {\bibfnamefont {D.~C.}\ \bibnamefont
  {McKay}}, \bibinfo {author} {\bibfnamefont {C.~J.}\ \bibnamefont {Wood}},
  \bibinfo {author} {\bibfnamefont {S.}~\bibnamefont {Sheldon}}, \bibinfo
  {author} {\bibfnamefont {J.~M.}\ \bibnamefont {Chow}}, \ and\ \bibinfo
  {author} {\bibfnamefont {J.~M.}\ \bibnamefont {Gambetta}},\ }\href {\doibase
  10.1103/PhysRevA.96.022330} {\bibfield  {journal} {\bibinfo  {journal} {Phys.
  Rev. A}\ }\textbf {\bibinfo {volume} {96}},\ \bibinfo {pages} {022330}
  (\bibinfo {year} {2017})}\BibitemShut {NoStop}%
\end{thebibliography}%

\clearpage
\appendix
\section{Device Setup}
	Our experiment is carried out on a 5-qubit superconducting quantum process, where two qubits are mainly manipulated. The device parameters are summarized in Table \uppercase\expandafter{\romannumeral2}.	
	\begin{table}[h]
			\begin{tabular}{c |c   c }
				\hline
				\hline
				&Q$_1$  &Q$_2$ \\
				\hline
				$\omega_{\textrm{r}}/2\pi$  (GHz)
				&6.403	&6.477 \\
				$\omega_{\textrm{q}}^{\textrm{0}}/2\pi$  (GHz)
				&5.299  &5.211	  \\
				$\omega_{\textrm{q}}/2\pi $ (GHz)
				&5.077  &4.889	\\
				$T_1$ ($\mu$s)
				&20.56  &26.32   \\
				$T_2^*$ ($\mu$s)
				& 2.52  &2.16	  \\
				$\chi_{qr}/2\pi$ (MHz)
				&1.05	&0.85	\\  			
				$U/2\pi $ (MHz)
				&$-235$ &$-235$ \\
				$f_{00}$
				&0.993	&0.996	\\
				$f_{11}$
				&0.966	&0.974 \\
				\hline
				\hline
			\end{tabular}
			
		\caption{
			{Device parameters: $\omega_{\textrm{r}}/2\pi$ is the frequency of readout resonator;
				$\omega_{\textrm{q}}^{\textrm{0}}/2\pi$ is the maximum frequency; $\omega_{\textrm{q}}/2\pi$s is the idle
				frequency; $T_1$ is the energy relaxation time of qubit measured at the idle frequency; $T_2^*$ is the dephasing time of
				qubit measured at the idle frequency; $\chi_{qr}$ is the dispersive shift; $U$ is the anharmonicity of qubit measured at the idle frequency; $f_{11}$ ($f_{00}$) is the readout fidelity of $|1\rangle$ ($|0\rangle$), the rate of correctly measuring $|1\rangle$ ($|0\rangle$) when the qubit is prepared at $|1\rangle$ ($|0\rangle$).}
		}
	\end{table}
	
\section{Experimental Setup}
	\begin{figure}[h]
	\centering
	\includegraphics[width=\linewidth]{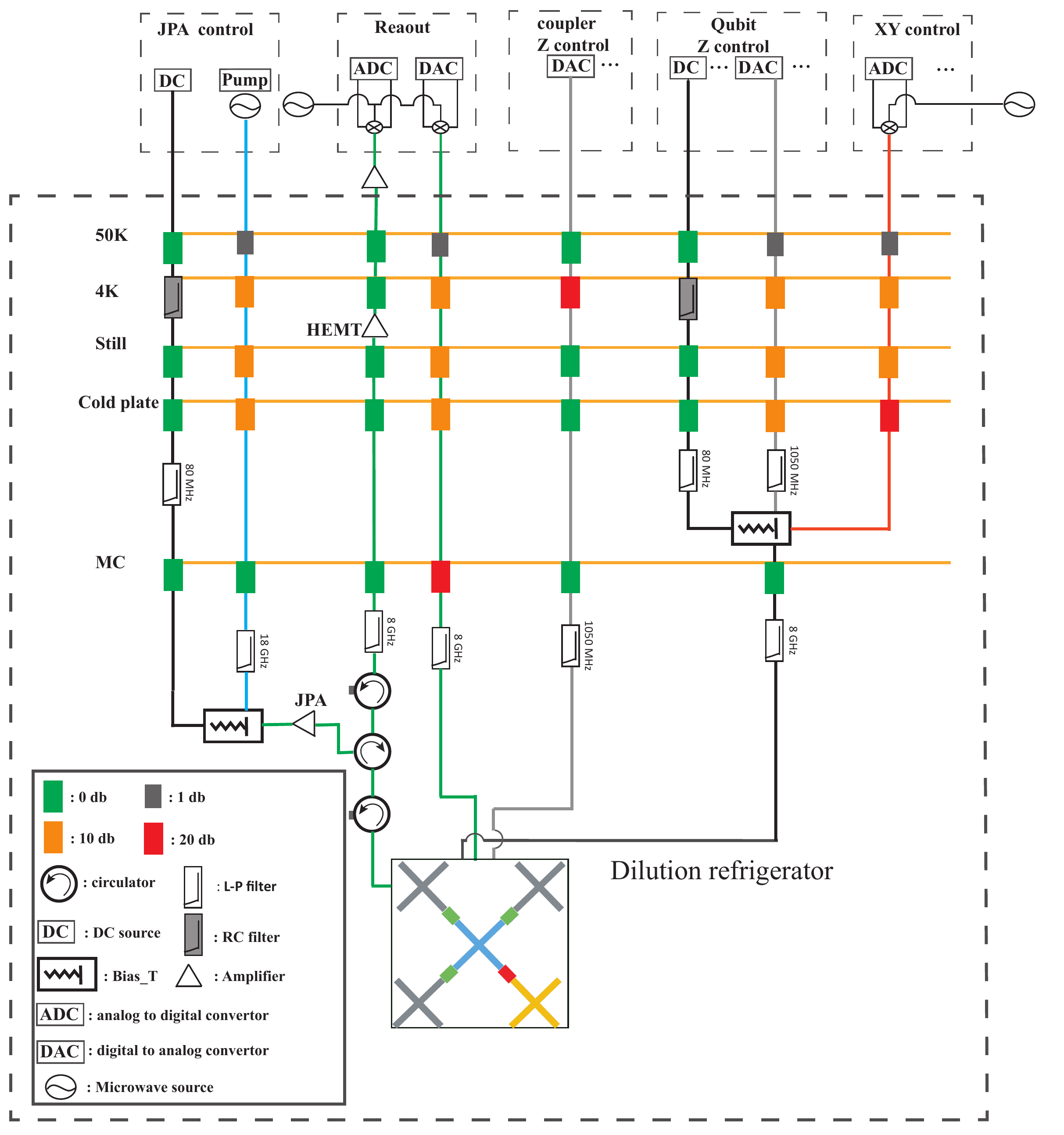}	
	\caption{The schematic diagram of control electronics and wiring.}
	\label{wiresetup}
	\end{figure}

	As shown in Fig.~\ref{wiresetup}, our measurement system consists of a dilution fridge, control electronics and wiring. The quantum processor which consists of two chips is installed at the base temperature stage of the dilution fridge. There are 41~dB  attenuation for XY control lines and 51~dB for readout lines at all stages in the dilution fridge. For qubit fast Z control lines, there is 31~dB attenuation at all stages, and 1050~MHz low pass filters are installed at cold plate stage. For coupler fast Z control lines, we need wider detune range so we attach 20~dB attenuation at all stages and 1050~MHz low pass filters at the base temperature stage. For qubit Z dc control lines, we have RC filters of 10~KHz cut-off frequency installed at 4~K stage and  80~MHz filters at cold plate stage. Qubit fast Z control line, dc Z control line and XY control line of are combined together with a bias T at cold plate stage. After three kinds of lines being combined together, we use 8~GHz low pass filter to lower high frequency noise. The readout signal firstly passes through a 8~GHz low pass filter and two circulators, then amplified by JPA. The third circulator is attached also at the base temperature and all the three circulators are used to block noise from higher temperature stages. Next, the readout signal is amplified by a high electron mobility transistor(HEMT) amplifier at 4~K stage, and then further amplified by a room temperature amplifier. Finally, signal is demodulated and digitized to extract the qubit-state information.
	
\end{document}